\documentclass[journal]{IEEEtran}
\usepackage{amsmath,graphicx,url,times,hyperref}
\usepackage{booktabs}
\usepackage[T1]{fontenc}

\usepackage[usenames, dvipsnames]{color}
\usepackage{booktabs}
\newcommand{\qk}[1] {{\color{black} #1}}

\usepackage{multirow}
\ifCLASSINFOpdf
\else
\fi
\begin{document}
%
% paper title
% Titles are generally capitalized except for words such as a, an, and, as,
% at, but, by, for, in, nor, of, on, or, the, to and up, which are usually
% not capitalized unless they are the first or last word of the title.
% Linebreaks \\ can be used within to get better formatting as desired.
% Do not put math or special symbols in the title.
\title{{\huge PANNs: Large-Scale Pretrained Audio Neural Networks for Audio Pattern Recognition}}
%
%
% author names and IEEE memberships
% note positions of commas and nonbreaking spaces ( ~ ) LaTeX will not break
% a structure at a ~ so this keeps an author's name from being broken across
% two lines.
% use \thanks{} to gain access to the first footnote area
% a separate \thanks must be used for each paragraph as LaTeX2e's \thanks
% was not built to handle multiple paragraphs
%

%\author{Michael~Shell,~\IEEEmembership{Member,~IEEE,}
%        John~Doe,~\IEEEmembership{Fellow,~OSA,}
%        and~Jane~Doe,~\IEEEmembership{Life~Fellow,~IEEE}% <-this % stops a space
%\thanks{M. Shell was with the Department
%of Electrical and Computer Engineering, Georgia Institute of Technology, %Atlanta,
%GA, 30332 USA e-mail: (see http://www.michaelshell.org/contact.html).}% <-this % stops a space
%\thanks{J. Doe and J. Doe are with Anonymous University.}% <-this % stops a space
%\thanks{Manuscript received April 19, 2005; revised August 26, 2015.}}

\author{Qiuqiang Kong,~\IEEEmembership{Student Member,~IEEE}, Yin Cao,~\IEEEmembership{Member,~IEEE}, Turab Iqbal, \\Yuxuan Wang, Wenwu Wang,~\IEEEmembership{Senior Member,~IEEE} and Mark D. Plumbley,~\IEEEmembership{Fellow,~IEEE}\thanks{Q. Kong, Y. Cao, T. Iqbal, and M. D. Plumbley are with the Centre for Vision, Speech and Signal Processing, University of Surrey, Guildford GU2 7XH, U.K. (e-mail: q.kong@surrey.ac.uk; yin.cao@surrey.ac.uk;  t.iqbal@surrey.ac.uk; m.plumbley@surrey.ac.uk).}
\thanks{This work was supported in part by the EPSRC Grant EP/N014111/1 ``Making Sense of Sounds'', in part by the Research Scholarship from the China Scholarship Council 201406150082, and in part by a studentship (Reference: 1976218) from the EPSRC Doctoral Training Partnership under Grant EP/N509772/1. This work was supported by National Natural Science Foundation of China (Grant No. 11804365). \textit{(Qiuqiang Kong is first author.) (Yin Cao is corresponding author.)}}
\thanks{Y. Wang is with the ByteDance AI Lab, Mountain View, CA, USA (e-mail:
wangyuxuan.11@bytedance.com).}
\thanks{W. Wang is with the Centre for Vision, Speech and Signal Processing,
University of Surrey, Guildford GU2 7XH, U.K., and also with Qingdao
University of Science and Technology, Qingdao 266071, China (e-mail:
w.wang@surrey.ac.uk).}}

\maketitle

% As a general rule, do not put math, special symbols or citations
% in the abstract or keywords.
\begin{abstract}
Audio pattern recognition is an important research topic in the machine learning area, and includes several tasks such as audio tagging, acoustic scene classification, music classification, speech emotion classification and sound event detection. Recently, neural networks have been applied to tackle audio pattern recognition problems. However, previous systems are built on specific datasets with limited durations. Recently, in computer vision and natural language processing, systems pretrained on large-scale datasets have generalized well to several tasks. However, there is limited research on pretraining systems on large-scale datasets for audio pattern recognition. In this paper, we propose pretrained audio neural networks (PANNs) trained on the large-scale AudioSet dataset. These PANNs are transferred to other audio related tasks. We investigate the performance and computational complexity of PANNs modeled by a variety of convolutional neural networks. We propose an architecture called Wavegram-Logmel-CNN using both log-mel spectrogram and waveform as input feature. Our best PANN system achieves a state-of-the-art mean average precision (mAP) of 0.439 on AudioSet tagging, outperforming the best previous system of 0.392. We transfer PANNs to six audio pattern recognition tasks, and demonstrate state-of-the-art performance in several of those tasks. We have released the source code and pretrained models of PANNs: \url{https://github.com/qiuqiangkong/audioset_tagging_cnn}.
\end{abstract}

% Note that keywords are not normally used for peer-review papers.

\section{Introduction}
Audio pattern recognition is an important research topic in the machine learning area, and plays an important role in our life. We are surrounded by sounds that contain rich information of where we are, and what events are happening around us. Audio pattern recognition contains several tasks such as audio tagging \cite{gemmeke2017audio}, acoustic scene classification \cite{mesaros2018multi}, music classification \cite{choi2016automatic}, speech emotion classification and sound event detection \cite{cakir2015polyphonic}. 

Audio pattern recognition has attracted increasing research interest in recent years. Early audio pattern recognition work focused on private datasets collected by individual researchers \cite{woodard1992modeling}\cite{ellis2001detecting}. For example, Woodard \cite{woodard1992modeling} applied a hidden Markov model (HMM) to classify three types of sounds: wooden door open and shut, dropped metal and poured water. Recently, the Detection and Classification of Acoustic Scenes and Events (DCASE) challenge series \cite{stowell2015detection}\cite{mesaros2018detection}\cite{mesaros2017dcase}\cite{mesaros2018multi} have provided publicly available datasets, such as acoustic scene classification and sound event detection datasets. The DCASE challenges have attracted increasing research interest in audio pattern recognition. For example, the recent DCASE 2019 challenge received 311 entries across five subtasks \cite{dcase2019_challenge}.

However, it is still an open question how well an audio pattern recognition system can perform when trained on large-scale datasets. In computer vision, several image classification systems have been built with the large-scale ImageNet dataset \cite{deng2009imagenet}. In natural language processing, several language models have been built with the large-scale text datasets such as Wikipedia \cite{devlin2018bert}. However, systems trained on large-scale audio datasets have been more limited \cite{gemmeke2017audio}\cite{hershey2017cnn}\cite{choi2017transfer}\cite{pons2017end}.

A milestone for audio pattern recognition was the release of AudioSet \cite{gemmeke2017audio}, a dataset containing over 5,000 hours of audio recordings with 527 sound classes. Instead of releasing the raw audio recordings, AudioSet released embedding features of audio clips extracted from a pretrained convolutional neural network \cite{hershey2017cnn}. Several researchers have investigated building systems with those embedding features \cite{hershey2017cnn}\cite{kong2018audio}\cite{yu2018multi}\cite{chou2018learning}\cite{wang2019comparison}\cite{kong2019weakly}. However, the embedding features may not be an optimal representation for audio recordings, which may limit the performance of those systems. In this article, we propose pretrained audio neural networks (PANNs) trained on raw AudioSet recordings with a wide range of neural networks. We show that several PANN systems outperform previous state-of-the-art audio tagging systems. We also investigate the audio tagging performance and computation complexities of PANNs. 

We propose that PANNs can be transferred to other audio pattern recognition tasks. Previous researchers have previously investigated transfer learning for audio tagging. For example, audio tagging systems were pretrained on the Million Song Dataset were proposed in \cite{van2014transfer}, with embedding features extracted from pretrained convolutional neural networks (CNNs) are used as inputs to second-stage classifiers such as neural networks or support vector machines (SVMs) \cite{choi2017transfer}\cite{wang2018polyphonic}. Systems pretrained on MagnaTagATune \cite{law2009input} and acoustic scene \cite{mesaros2016tut} datasets were fine-tuned on other audio tagging tasks \cite{pons2019musicnn}\cite{diment2017transfer}. These transfer learning systems were mainly trained with music datasets, and were limited to smaller datasets than AudioSet. 

\qk{The contribution of this work includes: (1) We introduce PANNs trained on AudioSet with 1.9 million audio clips with an ontology of 527 sound classes; (2) We investigate the trade-off between audio tagging performance and computation complexity of a wide range of PANNs; (3) We propose a system that we call Wavegram-Logmel-CNN that achieves a mean average precision (mAP) of 0.439 on AudioSet tagging, outperforming previous state-of-the-art system with an mAP 0.392 and Google's system with an mAP 0.314; (4) We show that PANNs can be transferred to other audio pattern recognition tasks, outperforming several state-of-the-art systems; (5) We have released the source code and pretrained PANN models.}

This paper is organized as follows: Section \ref{section:at_systems} introduces audio tagging with various convolutional neural networks; Section \ref{section:wavegram} introduces our proposed Wavegram-CNN systems; Section \ref{section:data_processing} introduces our data processing techniques, including data balancing and data augmentation for AudioSet tagging; Section \ref{section:experiments} shows experimental results, and Section \ref{section:conclusion} concludes this work.

\IEEEpeerreviewmaketitle

\section{Audio tagging systems}\label{section:at_systems}
Audio tagging is an essential task of audio pattern recognition, with the goal of predicting the presence or absence of audio tags in an audio clip. Early work on audio tagging included using manually-designed features as input, such as audio energy, zero-crossing rate, and mel-frequency cepstrum coefficients (MFCCs) \cite{li2001classification}. Generative models, including Gaussian mixture models (GMMs) \cite{vuegen2013mfcc}\cite{mesaros2010acoustic}, hidden Markov models (HMMs), and discriminative support vector machines (SVMs) \cite{uzkent2012non} have been used as classifiers. Recently, neural network based methods such as convolutional neural networks (CNNs) have been used \cite{choi2016automatic} to predict the tags of audio recordings. CNN-based systems have achieved state-of-the-art performance in several DCASE challenge tasks including acoustic scene classification \cite{mesaros2018multi} and sound event detection \cite{cakir2015polyphonic}. However, many of those works focused on particular tasks with a limited number of sound classes, and were not designed to recognize a wide range of sound classes. In this article, we focus on training large-scale PANNs on AudioSet \cite{gemmeke2017audio} to tackle the general audio tagging problem.

\subsection{CNNs}\label{subsection:cnns}
\subsubsection{Conventional CNNs}
CNNs have been successfully applied to computer vision tasks such as image classification \cite{dai2017very}\cite{he2016deep}. A CNN consists of several convolutional layers. Each convolutional layer contains several kernels that are convolved with the input feature maps to capture their local patterns.  CNNs adopted for audio tagging \cite{choi2016automatic}\cite{kong2019weakly} often use log mel spectrograms as input \cite{choi2016automatic}\cite{kong2019weakly}. Short time Fourier transforms (STFTs) are applied to time-domain waveforms to calculate spectrograms. Then, mel filter banks are applied to the spectrograms, followed by a logarithmic operation to extract log mel spectrograms \cite{choi2016automatic}\cite{kong2019weakly}.

\begin{table}

\centering
\caption{CNNs for AudioSet tagging}
\resizebox{\columnwidth}{!}{%
\begin{tabular}{cccc} 
\hline
\multicolumn{1}{c|}{VGGish \cite{gemmeke2017audio}} & \multicolumn{1}{c|}{CNN6} & \multicolumn{1}{c|}{CNN10} & \multicolumn{1}{c}{CNN14}  \\ 

\hline
\multicolumn{1}{c|}{$ \begin{matrix} \text{Log-mel spectrogram} \\ 96 \ \text{frames} \times 64 \ \text{mel bins} \end{matrix}$} & \multicolumn{3}{c}{$ \begin{matrix} \text{Log-mel spectrogram} \\ 1000 \ \text{frames} \times 64 \ \text{mel bins} \end{matrix}$}  \\
\hline
\multicolumn{1}{c|}{$ \begin{matrix} 3 \times 3 \ @ \ 64 \\ \text{ReLU} \end{matrix} $} & \multicolumn{1}{c|}{$ \begin{matrix} 5 \times 5 \ @ \ 64 \\ \text{BN, ReLU} \end{matrix} $} & \multicolumn{1}{c|}{$ \begin{pmatrix} 3 \times 3 \ @ \ 64 \\ \text{BN, ReLU} \end{pmatrix} \times 2 $} & $ \begin{pmatrix} 3 \times 3 \ @ \ 64 \\ \text{BN, ReLU} \end{pmatrix} \times 2 $  \\ 
\hline
\multicolumn{1}{c|}{MP $ 2 \times 2 $} & \multicolumn{3}{c}{Pooling $ 2 \times 2 $}  \\
\hline
\multicolumn{1}{c|}{$ \begin{matrix} 3 \times 3 \ @ \ 128 \\ \text{ReLU} \end{matrix} $} & \multicolumn{1}{l|}{$ \begin{matrix} 5 \times 5 \ @ \ 128 \\ \text{BN, ReLU} \end{matrix} $} & \multicolumn{1}{c|}{$ \begin{pmatrix} 3 \times 3 \ @ \ 128 \\ \text{BN, ReLU} \end{pmatrix} \times 2 $} & $ \begin{pmatrix} 3 \times 3 \ @ \ 128 \\ \text{BN, ReLU} \end{pmatrix} \times 2 $  \\ 
\hline
\multicolumn{1}{c|}{MP $ 2 \times 2 $} & \multicolumn{3}{c}{Pooling $ 2 \times 2 $}  \\
\hline
\multicolumn{1}{c|}{$ \begin{pmatrix} 3 \times 3 \ @ \ 256 \\ \text{ReLU} \end{pmatrix} \times 2 $} & \multicolumn{1}{c|}{$ \begin{matrix} 5 \times 5 \ @ \ 256 \\ \text{BN, ReLU} \end{matrix} $} & \multicolumn{1}{c|}{$ \begin{pmatrix} 3 \times 3 \ @ \ 256 \\ \text{BN, ReLU} \end{pmatrix} \times 2 $} & $ \begin{pmatrix} 3 \times 3 \ @ \ 256 \\ \text{BN, ReLU} \end{pmatrix} \times 2 $  \\ 
\hline
\multicolumn{1}{c|}{MP $ 2 \times 2 $} & \multicolumn{3}{c}{Pooling $ 2 \times 2 $}  \\
\hline
\multicolumn{1}{c|}{$ \begin{pmatrix} 3 \times 3 \ @ \ 512 \\ \text{ReLU} \end{pmatrix} \times 2 $} & \multicolumn{1}{c|}{$ \begin{matrix} 5 \times 5 \ @ \ 512 \\ \text{BN, ReLU} \end{matrix} $} & \multicolumn{1}{c|}{$ \begin{pmatrix} 3 \times 3 \ @ \ 512 \\ \text{BN, ReLU} \end{pmatrix} \times 2 $} & $ \begin{pmatrix} 3 \times 3 \ @ \ 512 \\ \text{BN, ReLU} \end{pmatrix} \times 2 $  \\ 
\hline
\multicolumn{1}{c|}{$ \begin{matrix} \text{MP} \ 2 \times 2 \\ \text{Flatten} \end{matrix} $} & \multicolumn{2}{c|}{Global pooling} & \multicolumn{1}{c}{Pooling $ 2 \times 2 $} \\
\hline
\multicolumn{1}{c|}{$ \begin{matrix} \text{FC} \ 4096 \\ \text{ReLU} \end{matrix} \times 2 $} & \multicolumn{2}{c|}{FC 512, ReLU} & \multicolumn{1}{c}{$ \begin{pmatrix} 3 \times 3 \ @ \ 1024 \\ \text{BN, ReLU} \end{pmatrix} \times 2 $} \\
\hline
\multicolumn{1}{c|}{FC 527, Sigmoid} & \multicolumn{2}{c|}{FC 527, Sigmoid} & \multicolumn{1}{c}{Pooling $ 2 \times 2 $} \\
\hline
                      &                       &                       & \multicolumn{1}{|c}{$ \begin{pmatrix} 3 \times 3 \ @ \ 2048 \\ \text{BN, ReLU} \end{pmatrix} \times 2 $}  \\ 
\cline{4-4}
                      &                       &                       &  \multicolumn{1}{|c}{Global pooling} \\ 
\cline{4-4}
                      &                       &                       &  \multicolumn{1}{|c}{FC 2048, ReLU} \\ 
\cline{4-4}
                      &                       &                       &  \multicolumn{1}{|c}{FC 527, Sigmoid} \\
\cline{4-4}
\end{tabular}}
\label{table:cnn_architectures}
\end{table}

\subsubsection{Adapting CNNs for AudioSet tagging}
The PANNs we use are based on our previously-proposed cross-task CNN systems for the DCASE 2019 challenge \cite{kong2019cross}, with an extra fully-connected layer added to the penultimate layer of CNNs to further increase the representation ability. We investigate 6-, 10- and 14-layer CNNs. The 6-layer CNN consists of 4 convolutional layers with a kernel size of $ 5 \times 5 $, based on AlexNet \cite{krizhevsky2012imagenet}. The 10- and 14-layer CNNs consist of 4 and 6 convolutional layers, respectively, inspired by the VGG-like CNNs \cite{simonyan2014very}. Each convolutional block consists of 2 convolutional layers with a kernel size of $ 3 \times 3 $. Batch normalization \cite{ioffe2015batch} is applied between each convolutional layer, and the ReLU nonlinearity \cite{nair2010rectified} is used to speed up and stabilize the training. We apply average pooling of size of $ 2 \times 2 $ to each convolutional block for downsampling, as $ 2 \times 2 $ average pooling has been shown to outperform $ 2 \times 2 $ max pooling \cite{kong2019cross}.

Global pooling is applied after the last convolutional layer to summarize the feature maps into a fixed-length vector. In \cite{pons2017end}, maximum and average operation were used for global pooling. To combine their advantages, we sum the averaged and maximized vectors. In our previous work \cite{kong2019cross}, those fixed-length vectors were used as embedding features for audio clips. In this work, we add an extra fully-connected layer to the fixed length vectors to extract embedding features which can further increase their representation ability. For a particular audio pattern recognition task, a linear classifier is applied to the embedding features, followed by either a softmax nonlinearity for classification tasks or a sigmoid nonlinearity for tagging tasks. Dropout \cite{srivastava2014dropout} is applied after each downsampling operation and fully connected layers to prevent systems from overfitting. Table \ref{table:cnn_architectures} summarizes our proposed CNN systems. The number after the ``$ @ $'' symbol indicates the number of feature maps. The first column shows the VGGish network proposed by \cite{hershey2017cnn}. MP is the abbreviation of max pooling. The ``Pooling $ 2 \times 2 $'' in Table \ref{table:cnn_architectures} is average pooling with size of $ 2 \times 2 $. In \cite{hershey2017cnn}, an audio clip was split into 1-second segments, \cite{hershey2017cnn} also assumed each segment inherits the label of the audio clip, which may lead to incorrect labels. In contrast, our systems from the second to the fourth columns in Table \ref{table:cnn_architectures} applies an entire audio clip for training without cutting the audio clip into segments.

We denote the waveform of an audio clip as $ x_{n} $, where $ n $ is the index of audio clips, and $ f(x_{n}) \in [0, 1]^{K} $ is the output of a PANN representing the presence probabilities of $ K $ sound classes. The label of $ x_{n} $ is denoted as $ y_{n} \in \{0, 1\}^{K} $. A binary cross-entropy loss function $ l $ is used to train a PANN:
\begin{equation} \label{eq1}
l = - \sum_{n=1}^{N} (y_{n} \cdot \text{ln} f(x_{n}) + (1 - y_{n}) \cdot \text{ln} (1 - f(x_{n})),
\end{equation}
\noindent where $ N $ is the number of training clips in AudioSet. In training, the parameters of $ f(\cdot) $ are optimized by using gradient descent methods to minimize the loss function $ l $.

\begin{table}
\centering
\caption{ResNets for AudioSet tagging}
\resizebox{\columnwidth}{!}{%
\begin{tabular}{ccc} 
\hline
\multicolumn{1}{c|}{ResNet22} & \multicolumn{1}{c|}{ResNet38} & ResNet54  \\ 
\hline
\multicolumn{3}{c}{Log mel spectrogram $ 1000 $ frames $ \times 64 $ mel bins} \\ 
\hline
\multicolumn{3}{c}{$ \begin{pmatrix} 3 \times 3 \ @ \ 512, \text{BN}, \text{ReLU} \end{pmatrix} \times 2 $} \\ 
\hline
\multicolumn{3}{c}{Pooling $ 2 \times 2 $} \\ 
\hline
\multicolumn{1}{c|}{$ \begin{pmatrix} \text{BasicB} \ @ \ 64 \end{pmatrix} \times 2 $} & \multicolumn{1}{c|}{$ \begin{pmatrix} \text{BasicB} \ @ \ 64 \end{pmatrix} \times 3 $} & \multicolumn{1}{c}{$ \begin{pmatrix} \text{BottleneckB} \ @ \ 64 \end{pmatrix} \times 3 $}  \\ 
\hline
\multicolumn{3}{c}{Pooling $ 2 \times 2 $} \\
\hline
\multicolumn{1}{c|}{$ \begin{pmatrix} \text{BasicB} \ @ \ 128 \end{pmatrix} \times 2 $} & \multicolumn{1}{c|}{$ \begin{pmatrix} \text{BasicB} \ @ \ 128 \end{pmatrix} \times 4 $} & \multicolumn{1}{c}{$ \begin{pmatrix} \text{BottleneckB} \ @ \ 128 \end{pmatrix} \times 4 $}  \\ 
\hline
\multicolumn{3}{c}{Pooling $ 2 \times 2 $} \\
\hline
\multicolumn{1}{c|}{$ \begin{pmatrix} \text{BasicB} \ @ \ 256 \end{pmatrix} \times 2 $} & \multicolumn{1}{c|}{$ \begin{pmatrix} \text{BasicB} \ @ \ 256 \end{pmatrix} \times 6 $} & \multicolumn{1}{c}{$ \begin{pmatrix} \text{BottleneckB} \ @ \ 256 \end{pmatrix} \times 6 $}  \\ 
\hline
\multicolumn{3}{c}{Pooling $ 2 \times 2 $} \\
\hline
\multicolumn{1}{c|}{$ \begin{pmatrix} \text{BasicB} \ @ \ 512 \end{pmatrix} \times 2 $} & \multicolumn{1}{c|}{$ \begin{pmatrix} \text{BasicB} \ @ \ 512 \end{pmatrix} \times 3 $} & \multicolumn{1}{c}{$ \begin{pmatrix} \text{BottleneckB} \ @ \ 512 \end{pmatrix} \times 3 $}  \\ 
\hline
\multicolumn{3}{c}{Pooling $ 2 \times 2 $} \\
\hline
\multicolumn{3}{c}{$ \begin{pmatrix} 3 \times 3 \ @ \ 2048, \text{BN}, \text{ReLU} \end{pmatrix} \times 2 $}   \\ 
\hline
\multicolumn{3}{c}{Global pooling}  \\ 
\hline
\multicolumn{3}{c}{FC 2048, ReLU} \\ 
\hline
\multicolumn{3}{c}{FC 527, Sigmoid} \\
\hline
\end{tabular}}
\label{table:resnet_architecture}
\end{table}

\subsection{ResNets}
\subsubsection{Conventional residual networks (ResNets)}
Deeper CNNs have been shown to achieve better performance than shallower CNNs for audio classification \cite{dai2017very}. One challenge of very deep conventional CNNs is that the gradients do not propagate properly from the top layers to the bottom layers \cite{he2016deep}. To address this problem, ResNets \cite{he2016deep} introduced shortcut connections between convolutional layers. In this way, the forward and backward signals can be propagated from one layer to any other layer directly. The shortcut connections only introduce a small number of extra parameters and a little additional computational complexity. A ResNet consists of several blocks, where each block consists of two convolutional layers with a kernel size of $ 3 \times 3 $, and a shortcut connection between input and output. Each bottleneck block consists of three convolutional layers with a network-in-network architecture \cite{lin2013network} that can be used instead of the basic blocks in a ResNet \cite{he2016deep}.

\subsubsection{Adapting ResNets for AudioSet tagging}
We adapt ResNet \cite{he2016deep} for AudioSet tagging as follows. To begin with, two convolutional layers and a downsampling layer are applied on the log mel spectrogram to reduce the input log mel spectrogram size. We implement three types of ResNets with different depths: a 22-layer ResNet with 8 basic blocks; a 38-layer ResNet with 16 basic blocks, and a 54-layer ResNet with 16 residual blocks. Table \ref{table:resnet_architecture} shows the architecture of the ResNet systems adapted for AudioSet tagging. The BasicB and BottleneckB are abbreviations of basic block and bottleneck block, respectively. 

\begin{table}
\centering
\caption{MobileNets for AudioSet tagging}
\begin{tabular}{cc} 
\hline
\multicolumn{1}{c|}{MobileNetV1} & \multicolumn{1}{c}{MobileNetV2} \\ 
\hline
\multicolumn{2}{c}{$ 3 \times 3 \ @ \ 32 \text{, BN, ReLU} $} \\ 
\hline
\multicolumn{2}{c}{Pooling $ 2 \times 2 $} \\
\hline
\multicolumn{1}{c|}{$ \begin{matrix} \text{V1Block} \ @ \ 64 \\ \text{V1Block} \ @ \ 128 \end{matrix} $} & \multicolumn{1}{c}{$ \begin{matrix} \text{V2Block, t=1} \ @ \ 16 \\ \left ( \text{V2Block, t=6} \ @ \ 24 \right ) \times 2 \end{matrix} $} \\
\hline
\multicolumn{2}{c}{Pooling $ 2 \times 2 $} \\
\hline
\multicolumn{1}{c|}{$ \begin{matrix} \text{V1Block} \ @ \ 128 \\ \text{V1Block} \ @ \ 256 \end{matrix} $} & \multicolumn{1}{c}{$ \left ( \text{V2Block, t=6} \ @ \ 32 \right ) \times 3 $} \\
\hline
\multicolumn{2}{c}{Pooling $ 2 \times 2 $} \\
\hline
\multicolumn{1}{c|}{$ \begin{matrix} \text{V1Block} \ @ \ 256 \\ \text{V1Block} \ @ \ 512 \end{matrix} $} & \multicolumn{1}{c}{$ \left ( \text{V2Block, t=6} \ @ \ 64 \right ) \times 4 $} \\
\hline
\multicolumn{2}{c}{Pooling $ 2 \times 2 $} \\
\hline
\multicolumn{1}{c|}{$ \begin{matrix} \left ( \text{V1Block} \ @ \ 512 \right ) \times 5 \\ \text{V1Block} \ @ \ 1024 \end{matrix} $} & \multicolumn{1}{c}{$ \left ( \text{V2Block, t=6} \ @ \ 96 \right ) \times 3 $} \\
\hline
\multicolumn{2}{c}{Pooling $ 2 \times 2 $} \\
\hline
\multicolumn{1}{c|}{$ \text{V1Block} \ @ \ 1024 $} & \multicolumn{1}{c}{$ \begin{matrix} \left ( \text{V2Block, t=6} \ @ \ 160 \right ) \times 3 \\ \left ( \text{V2Block, t=6} \ @ \ 320 \right ) \times 1 \end{matrix} $} \\
\hline
\multicolumn{2}{c}{Global pooling} \\
\hline
\multicolumn{2}{c}{FC, 1024, ReLU} \\
\hline
\multicolumn{2}{c}{FC, 527, Sigmoid} \\
\hline
\end{tabular}
\label{table:mobilenet_architecture}
\end{table}

\subsection{MobileNets}\label{section:mobile_net}
\subsubsection{Conventional MobileNets}
Computational complexity is an important issue when systems are implemented on portable devices. Compared to CNNs and ResNets, MobileNets were intended to reduce the number of parameters and multiply-add operations in a CNN. MobileNets were based on depthwise separable convolutions \cite{howard2017mobilenets} by factorizing a standard convolution into a depthwise convolution and a $ 1 \times 1 $ pointwise convolution \cite{howard2017mobilenets}.

\subsubsection{Adapting MobileNets for AudioSet tagging}
We adapt MobileNetV1 \cite{howard2017mobilenets} and MobileNetV2 \cite{sandler2018mobilenetv2} systems for AudioSet tagging shown in Table \ref{table:mobilenet_architecture}. The V1Blocks and V2Blocks are MobileNet convolutional blocks \cite{howard2017mobilenets}\cite{sandler2018mobilenetv2}, each consisting of two and three convolutional layers, respectively.

\subsection{One-dimensional CNNs}

Previous audio tagging systems were based on the log mel spectrogram, a hand-crafted feature. To improve performance, several researchers proposed to build one-dimensional CNNs which operate directly on the time-domain waveforms. For example, Dai et al. \cite{dai2017very} proposed a one-dimensional CNN for acoustic scene classification, and Lee et al. \cite{lee2017sample} proposed a one-dimensional CNN that was later adopted by Pons et al. \cite{pons2017end} for music tagging.

\subsubsection{DaiNet}

DaiNet \cite{dai2017very} applied kernels of length 80 and stride 4 to the input waveform of audio recordings. The kernels are learnable during training. To begin with, a maximum operation is applied to the first convolutional layer, which is designed to make the system be robust to phase shift of the input signals. Then, several one-dimensional convolutional blocks with kernel size $ 3 $ and stride 4 were applied to extract high level features. An 18-layer DaiNet with four convolutional layers in each convolutional block achieved the best result in UrbanSound8K \cite{salamon2014dataset} classification \cite{dai2017very}. 

\subsubsection{LeeNet}

In contrast to DaiNet that applied large kernels in the first layer, LeeNet \cite{lee2017sample} applied small kernels with length 3 on the waveforms, to replace the STFT for spectrogram extraction. LeeNet consists of several one dimensional convolutional layers, each followed by a downsampling layer of size 2. The original LeeNet consists of 11 layers.

\subsubsection{Adapting one-dimensional CNNs for AudioSet tagging}\label{section:one_dim_cnns}
We modify LeeNet by extending it to a deeper architecture with 24 layers, replacing each convolutional layer with a convolutional block that consists of two convolutional layers. To further increase the number of layers of the one-dimensional CNNs, we propose a one-dimensional residual network (Res1dNet) with a small kernel size of 3. We replace the convolutional blocks in LeeNet with residual blocks, where each residual block consists of two convolutional layers with kernel size 3. The first and second convolutional layers of convolutional block have dilations of 1 and 2, respectively, to increase the receptive field of the corresponding residual block. Downsampling is applied after each residual block. By using 14 and 24 residual blocks, we obtain a Res1dNet31 and a Res1dNet51 with 31 and 51 layers, respectively.

\section{Wavegram-CNN systems}\label{section:wavegram}

Previous one-dimensional CNN systems \cite{dai2017very}\cite{lee2017sample}\cite{pons2017end} have not outperformed the systems trained with log mel spectrograms as input. One characteristic of previous time-domain CNN systems \cite{dai2017very}\cite{lee2017sample} is that they were not designed to capture frequency information, because there is no frequency axis in the one-dimensional CNN systems, so they can not capture frequency patterns of a sound event with different pitch shifts.

\subsection{Wavegram-CNN systems}
\begin{figure}[t]
  \centering
  \centerline{\includegraphics[width=\columnwidth]{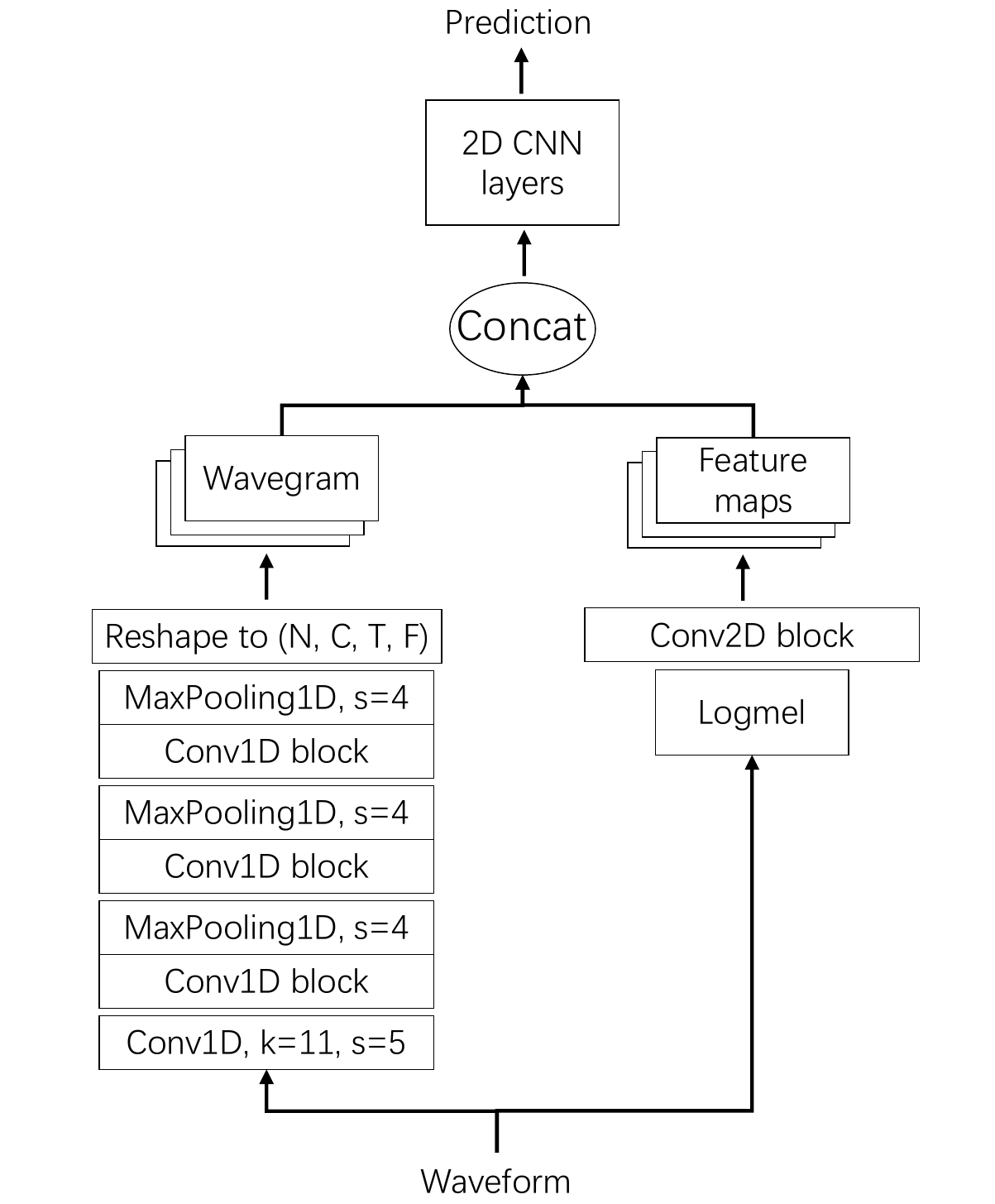}}
  \caption{Architecture of Wavegram-Logmel-CNN}
  \label{fig:wavegram_logmel_cnn}
\end{figure}

\qk{In this section, we propose architectures which we call Wavegram-CNN and Wavegram-Logmel-CNN for AudioSet tagging. The Wavegram-CNN we propose is a time-domain audio tagging system. Wavegram is a feature we propose that is similar to log mel spectrogram, but is learnt using a neural network. A Wavegram is designed to learn a time-frequency representation that is a modification of the Fourier transform. A Wavegram has a time axis and a frequency axis. Frequency patterns are important for audio pattern recognition, for example, sounds with different pitch shifts belong to the same class. A Wavegram is designed to learn frequency information that may be lacking in one-dimensional CNN systems. Wavegrams may also improve over hand-crafted log mel spectrograms by learning a new kind of time-frequency transform from data. Wavegrams can then replace log mel spectrograms as input features resulting in our Wavegram-CNN system. We also combine the Wavegram and the log mel spectrogram as a new feature to build the Wavegram-Logmel-CNN system as shown in Fig. \ref{fig:wavegram_logmel_cnn}.}

To build a Wavegram, we first apply a one-dimensional CNN to time-domain waveform. The one-dimensional CNN begins with a convolutional layer with filter length 11 and stride 5 to reduce the size of the input. This immediately reduces the input lengths by a factor of 5 times to reduce memory usage. This is followed by three convolutional blocks, where each convolutional block consists of two convolutional layers with dilations of 1 and 2, respectively, which are designed to increase the receptive field of the convolutional layers. Each convolutional block is followed by a downsampling layer with stride 4. By using the stride and downsampling three times, we downsample a 32 kHz audio recording to $32,000 / 5 / 4 / 4 / 4 = 100$ frames of features per second. We denote the output size of the one-dimensional CNN layers as $ T \times C $, where $ T $ is the number of frames and $ C $ is the number of channels. \qk{We reshape this output to a tensor with a size of $ T \times F \times C / F $ by splitting $ C $ channels into $ C / F $ groups, where each group has $ F $ frequency bins. We call this tensor a \textit{Wavegram}. The Wavegram learns frequency information by introducing $F$ frequency bins in each of $ C / F $ channels.} \qk{We apply CNN14 described in Section \ref{subsection:cnns} as a backbone architecture on the extracted Wavegram, so that we can fairly compare the Wavegram and log mel spectrogram based systems. Two dimensional CNNs such as CNN14 can capture time-frequency invariant patterns on the Wavegram, because kernels are convolved along both time and frequency axis in a Wavegram.}

\subsection{Wavegram-Logmel-CNN}
Furthermore, we can combine the Wavegram and log mel spectrogram into a new representation. In this way, we can utilize the information from both time-domain waveforms and log mel spectrograms. The combination is carried out along the channel dimension. The Wavegram provides extra information for audio tagging, complementing the log mel spectrogram. Fig. \ref{fig:wavegram_logmel_cnn} shows the architecture of the Wavegram-Logmel-CNN.

\section{Data processing}\label{section:data_processing}
In this section, we introduce data processing for AudioSet tagging, including data balancing and data augmentation. Data balancing is a technique used to train neural networks on a highly unbalanced dataset. Data augmentation is a technique used to augment the dataset, to prevent systems from overfitting during training. 

\subsection{Data balancing}\label{section:data_balancing}
The number of audio clips available for training varies from sound class to sound class. For example, there are over 900,000 audio clips belonging to the categories ``Speech'' and ``Music''. On the other hand, there are only tens of audio clips belonging to the category ``Toothbrush''. The number of audio clips of different sound classes has a long tailed distribution. Training data are input to a PANN in mini-batches during training. Without a data balancing strategy, audio clips are uniformly sampled from AudioSet. Therefore, sound classes with more training clips such as ``Speech'' are more likely to be sampled during training. In an extreme case, all data in a mini-batch may belong to the same sound class. This will cause the PANN to overfit to sound classes with more training clips, and underfit to sound classes with fewer training clips. \qk{To solve this problem, we design a balanced sampling strategy to train PANNs. That is, audio clips are approximately equally sampled from all sound classes to constitute a mini-batch.} We use the term ``approximately'' because an audio clip may contain more than one tag.

\subsection{Data augmentation}
Data augmentation is a useful way to prevent a system from overfitting. Some sound classes in AudioSet contain only a small number (e.g., hundreds) of training clips which may limit the performance of PANNs. We apply mixup \cite{zhang2017mixup} and SpecAugment \cite{park2019specaugment} to augment data during training. 

\subsubsection{Mixup}
Mixup \cite{zhang2017mixup} is a way to augment a dataset by interpolating both the input and target of two audio clips from a dataset. For example, we denote the input of two audio clips as $ x_{1}, x_{2} $, and their targets as $ y_{1}, y_{2} $, respectively. Then, the augmented input and target can be obtained by $ x = \lambda x_{1} + (1 - \lambda) x_{2} $ and $ y = \lambda y_{1} + (1 - \lambda) y_{2} $ respectively, where $ \lambda $ is sampled from a Beta distribution \cite{zhang2017mixup}. \qk{By default, we apply mixup on the log mel spectrogram. We will compare the performance of mixup augmentation on the log mel spectrogram and on the time-domain waveform in Section \ref{section:exp_augmentation}}.

\subsubsection{SpecAugment}
SpecAugment \cite{park2019specaugment} was proposed for augmenting speech data for speech recognition. SpecAugment operates on the log mel spectrogram of an audio clip using frequency masking and time masking. Frequency masking is applied such that $ f $ consecutive mel frequency bins $ [ f_{0}, f_{0} + f ] $ are masked, where $ f $ is chosen from a uniform distribution from 0 to a frequency mask parameter $ f' $, and $ f_{0} $ is chosen from $ [0, F - f] $, where $ F $ is the number of mel frequency bins \cite{park2019specaugment}. There can be more than one frequency mask in each log mel spectrogram. The frequency mask can improve the robustness of PANNs to frequency distortion of audio clips \cite{park2019specaugment}. Time masking is similar to frequency masking, but is applied in the time domain.

\begin{figure}[t]
  \centering
  \centerline{\includegraphics[width=\columnwidth]{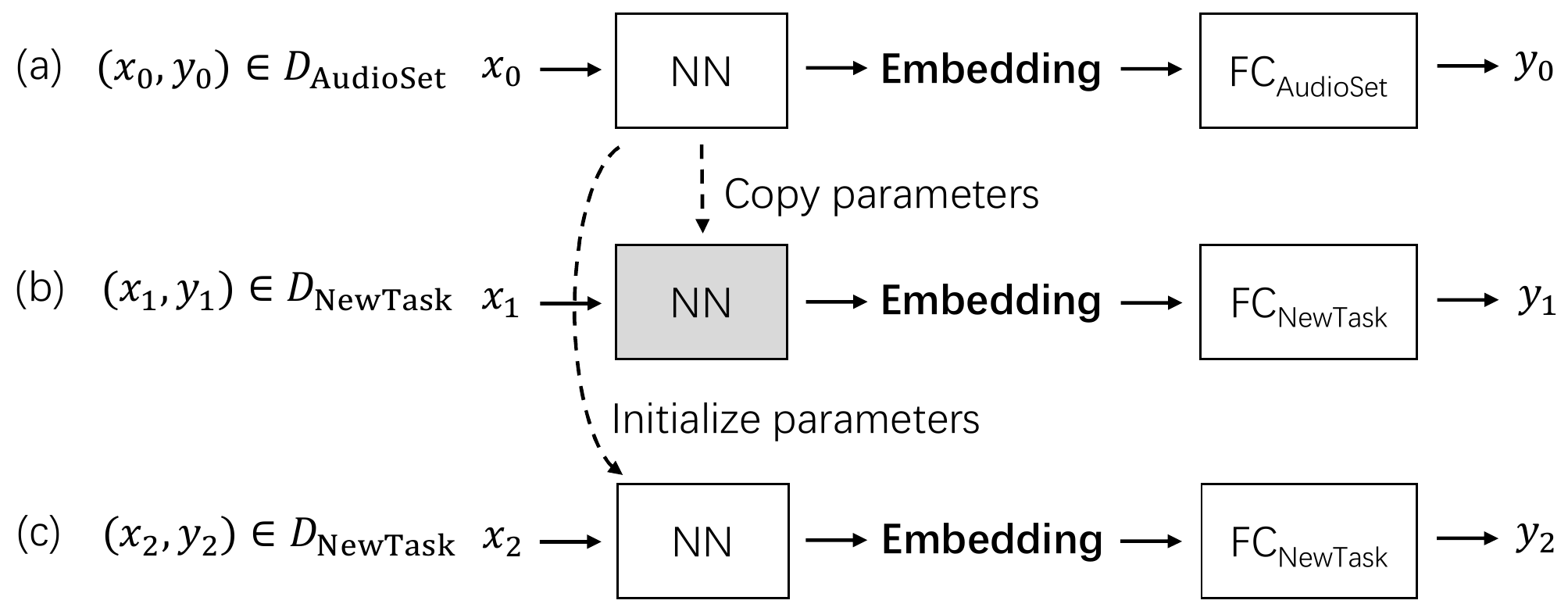}}
  \caption{(a) A PANN is pretrained with the AudioSet dataset. (b) For a new task, the PANN is used as a feature extractor. A classifier is built on the extracted embedding features. The shaded rectangle indicates the parameters are frozen and not trained. (c) For a new task, the parameters of a neural network are initialized with a PANN. Then, all parameters are fine-tuned on the new task.}
  \label{fig:transfer_framework}
\end{figure}

\begin{figure*}[t]
  \centering
  \centerline{\includegraphics[width=\textwidth]{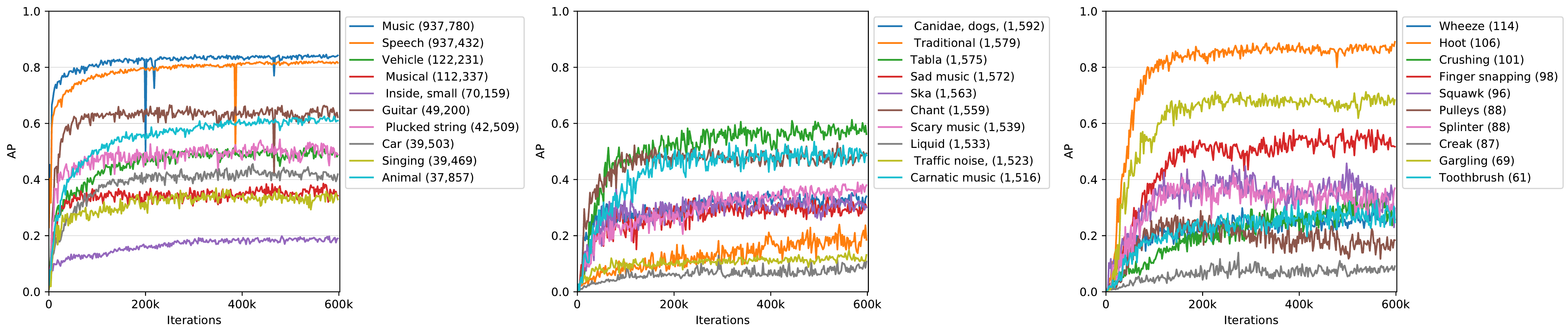}}
  \caption{Class-wise AP of sound events with the CNN14 system. The number inside parentheses indicates the number of training clips. The left, middle, right columns show the AP of sound classes with the number of training clips ranked the 1st to 10th, 250th to 260th and 517th to 527th in the training set of AudioSet.}
  \label{fig:class_iteration_map}
\end{figure*}

\section{Transfer to other tasks}\label{section:transfer}
To investigate the generalization ability of PANNs, we transfer PANNs to a range of audio pattern recognition tasks. Previous works on audio transfer learning \cite{van2014transfer}\cite{choi2017transfer}\cite{wang2018polyphonic}\cite{pons2019musicnn}\cite{law2009input} mainly focused on music tagging, and were limited to smaller datasets than AudioSet. To begin with, we demonstrate the training of a PANN in Fig. \ref{fig:transfer_framework}(a). Here, $ D_{\text{AudioSet}} $ is the AudioSet dataset, and $ x_{0} $, $ y_{0} $ are training input and target, respectively. $ \text{FC}_\text{AudioSet} $ is the fully connected layer for AudioSet tagging. In this article, we propose to compare the following transfer learning strategies. 

1) Train a system from scratch. All parameters are randomly initialized. Systems are similar to PANNs, except for the final fully-connected layer which depends on the task dependent number of outputs. This system is used as a baseline system to be compared with other transfer learning systems. 

2) Use a PANN as a feature extractor. For new tasks, the embedding features of audio clips are calculated by using the PANN. Then, the embedding features are used as input to a classifier, such as a fully-connected neural network. When training on new tasks, the parameters of the PANN are frozen and not trained. Only the parameters of the classifier built on the embedding features are trained. Fig. \ref{fig:transfer_framework}(b) shows this strategy, where $ D_{\text{NewTask}} $ is a new task dataset, and $ \text{FC}_\text{NewTask} $ is the fully connected layer of a new task. The PANN is used as a feature extractor. A classifier is built on the extracted embedding features. The shaded rectangle indicates the parameters which are frozen and not trained.

(3) Fine-tune a PANN. A PANN is used for a new task, except the final fully-connected layer. All parameters are initialized from the PANN, except the final fully-connected layer which is randomly initialized. All parameters are fine-tuned on $ D_{\text{NewTask}} $. Fig. \ref{fig:transfer_framework}(c) demonstrates the fine-tuning of a PANN.

\section{Experiments}\label{section:experiments}

First, we evaluate the performance of PANNs on AudioSet tagging. Then, the PANNs are transferred to several audio pattern recognition tasks, including acoustic scene classification, general audio tagging, music classification and speech emotion classification. 

\subsection{AudioSet dataset}
AudioSet is a large-scale audio dataset with an ontology of 527 sound classes \cite{gemmeke2017audio}. The audio clips from AudioSet are extracted from YouTube videos. The training set consists of 2,063,839 audio clips, including a ``balanced subset'' of 22,160 audio clips, where there are at least 50 audio clips for each sound class. The evaluation set consists of 20,371 audio clips. Instead of using the embedding features provided by \cite{gemmeke2017audio}, we downloaded raw audio waveforms of AudioSet in Dec. 2018 using the links provided by \cite{gemmeke2017audio}, and ignored the audio clips that are no longer downloadable. We successfully download 1,934,187 (94\%) of the audio clips of the full training set, including 20,550 (93\%) of the audio clips of the balanced training set. We successfully download 18,887 audio clips of the evaluation dataset. We pad the audio clips to 10 seconds with silence if they are shorter than 10 seconds. Considering the fact that a large number of audio clips from YouTube are monophonic and have a low sampling rate, we convert all audio clips to monophonic and resample them to 32 kHz. 

For the CNN systems based on log mel spectrograms, STFT is applied on the waveforms with a Hamming window of size 1024 \cite{kong2019cross} and a hop size of 320 samples. This configuration leads to 100 frames per second. Following \cite{kong2019cross}, we apply 64 mel filter banks to calculate the log mel spectrogram. The lower and upper cut-off frequencies of the mel banks are set to 50 Hz and 14 kHz to remove low frequency noise and the aliasing effects. We use the \textit{torchlibrosa}\footnote{\url{https://github.com/qiuqiangkong/torchlibrosa}}, a PyTorch implementation of functions of librosa \cite{mcfee2015librosa} to build log mel spectrogram extraction into PANNs. The log mel spectrogram of a 10-second audio clip has a shape of $ 1001 \times 64 $. The extra one frame is caused by 
applying the ``centre'' argument when calculating STFT. A batch size of 32, and an Adam \cite{kingma2014adam} optimizer with a learning rate of 0.001 is used for training. Systems are trained on a single card Tesla-V100-PCIE-32GB. Each system takes around 3 days to train from scratch for 600 k iterations.

\begin{table}[t]
\centering
\caption{Comparison with previous methods}
\label{table:comparison}
\begin{tabular}{*{4}{c}}
 \toprule
 & mAP & AUC & d-prime \\
 \midrule
 Random guess & 0.005 & 0.500 & 0.000 \\
 Google CNN \cite{gemmeke2017audio} & 0.314 & 0.959 & 2.452 \\
 Single-level attention \cite{kong2018audio} & 0.337 & 0.968 & 2.612 \\
 Multi-level attention \cite{yu2018multi} & 0.360 & 0.970 & 2.660 \\
 Large feature-level attention \cite{kong2019weakly} & 0.369 & 0.969 & 2.640 \\
 TAL Net \cite{wang2019comparison} & 0.362 & 0.965 & 2.554 \\
 DeepRes \cite{ford2019deep} & 0.392 & 0.971 & 2.682 \\
 \midrule
 Our proposed CNN14 & \textbf{0.431} & \textbf{0.973} & \textbf{2.732} \\
 \bottomrule
\end{tabular}
\end{table}

\subsection{Evaluation metrics}
Mean average precision (mAP), mean area under the curve (mAUC) and d-prime are used as official evaluation metrics for AudioSet tagging \cite{kong2019weakly}\cite{gemmeke2017audio}. Average precision (AP) is the area under the recall and precision curve. AP does not depend on the number of true negatives, because neither precision nor recall depends on the number of true negatives. On the other hand, AUC is the area under the false positive rate and true positive rate (recall) which reflects the influence of the true negatives. The d-prime \cite{gemmeke2017audio} is also used as an metric and be calculated directly from AUC \cite{gemmeke2017audio}. All metrics are calculated on individual classes, and then averaged across all classes. Those metrics are also called macro metrics.

\begin{figure*}[t]
  \centering
  \centerline{\includegraphics[width=\textwidth]{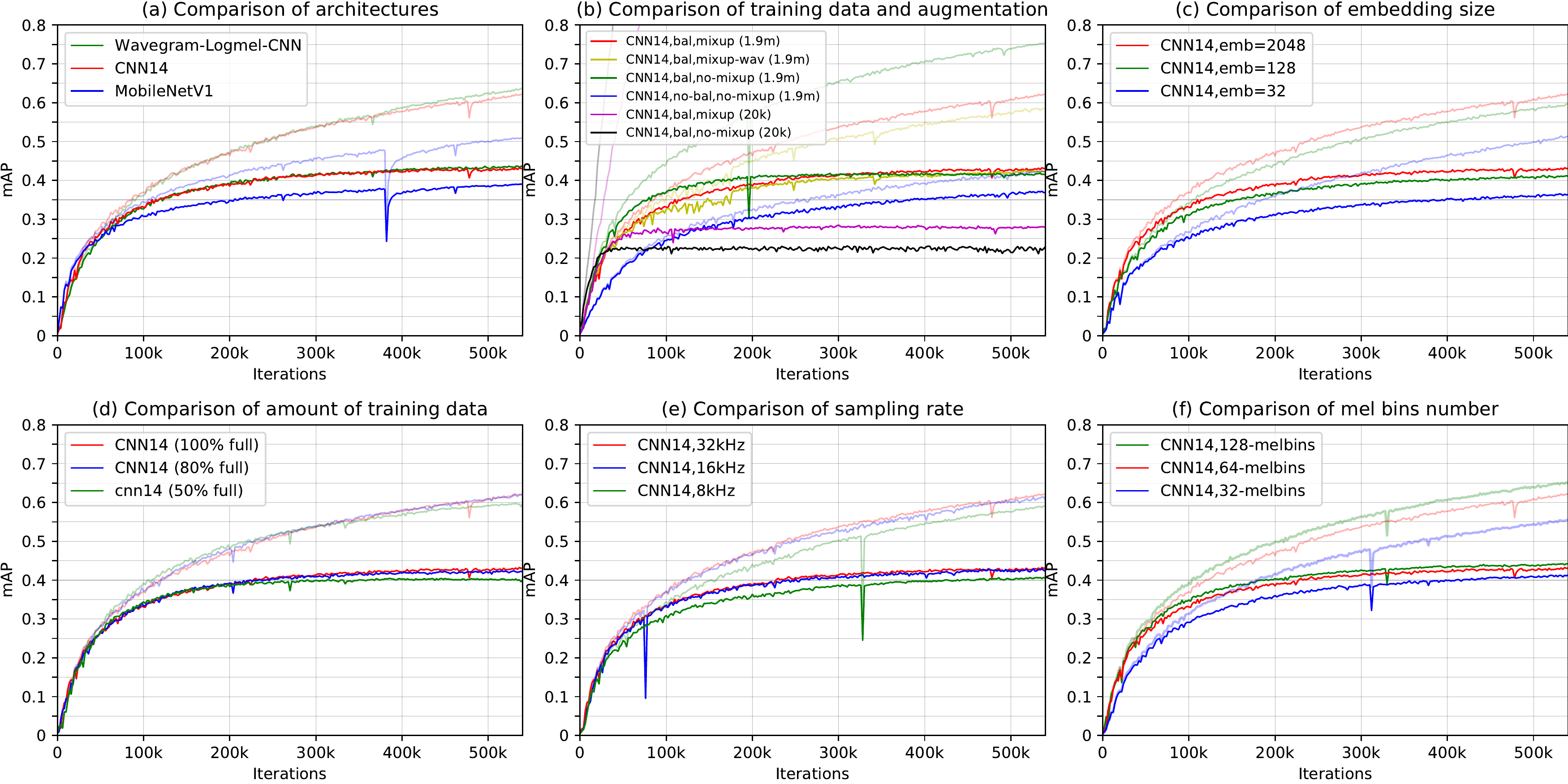}}
  \caption{Results of PANNs on AudioSet tagging. The transparent and solid lines are training mAP and evaluation mAP, respectively. The six plots show the results with different: (a) architectures; (b) data balancing and data augmentation; (c) embedding size; (d) amount of training data; (e) sampling rate; (f) number of mel bins.}
  \label{fig:six_figures}
\end{figure*}

\subsection{AudioSet tagging results}
\subsubsection{Comparison with previous methods}
Table \ref{table:comparison} shows the comparison of our proposed CNN14 system with previous AudioSet tagging systems. We choose CNN14 as a basic model to investigate a various of hyper-parameter configurations for AudioSet tagging, because CNN14 is a standard CNN that has a simple architecture, and can be compared with previous CNN systems \cite{choi2016automatic}\cite{kong2019cross}. As a baseline, random guess achieves an mAP of 0.005, an AUC of 0.500 and a d-prime of 0.000, respectively. The result released by Google \cite{gemmeke2017audio} trained with embedding features from \cite{hershey2017cnn} achieved an mAP of 0.314 and an AUC of 0.959, respectively. The single-level attention and multi-level attention systems \cite{kong2018audio, yu2018multi} achieved mAPs of 0.337 and 0.360, which were later improved by a feature-level attention neural network that achieved an mAP of 0.369. Wang et al. \cite{wang2019comparison} investigated five different types of attention functions and achieved an mAP of 0.362. All the above systems were built on the embedding features released with AudioSet \cite{gemmeke2017audio}. The recent DeepRes system \cite{ford2019deep} was built on waveforms downloaded from YouTube, and achieved an mAP of 0.392. The bottom rows of Table \ref{table:comparison} shows our proposed CNN14 system achieves an mAP of 0.431, outperforming the best of previous systems. We use CNN14 as a backbone to build Wavegram-Logmel-CNN for fair comparison with the CNN14 system. Fig. \ref{fig:six_figures}(a) shows that Wavegram-Logmel-CNN outperforms the CNN14 system, and the MobileNetV1 system. Detailed results are shown in later this section in Table \ref{table:all_systems}.

\subsubsection{Class-wise performance}
\begin{figure*}[t!]
  \centering
  \centerline{\includegraphics[width=\textwidth]{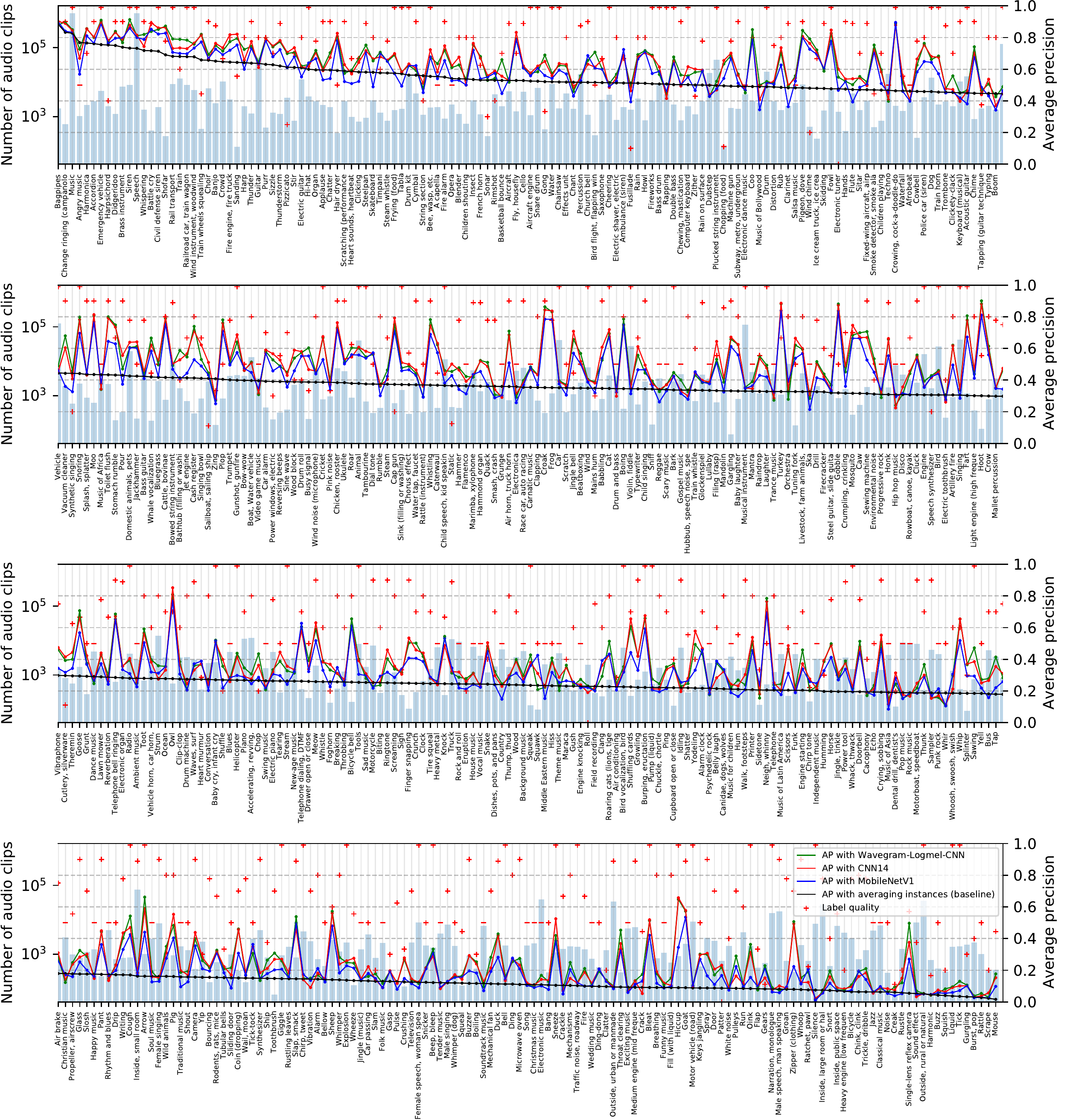}}
  \caption{Class-wise performance of AudioSet tagging systems. Red, blue and black curves are APs of CNN14, MobileNetV1 and the audio tagging system \cite{kong2019weakly}. The blue bars show the number of training clips in logarithmic scale.}
  \label{fig:long_fig}
\end{figure*}

Fig. \ref{fig:class_iteration_map} shows the class-wise AP of different sound classes with the CNN14 system. The left, middle, right columns show the AP of sound classes with the number of training clips ranked the 1st to 10th, 250th to 260th and 517th to 527th in the training set of AudioSet. The performance of different sound classes can be very different. For example, ``Music'' and ``Speech'' achieve APs of over 0.80. On the other hand, some sound classes such as ``Inside, small'' achieve an AP of only 0.19. Fig. \ref{fig:class_iteration_map} shows that APs are usually not correlated with the number of training clips. For example, the left column shows that ``Inside, small'' contains 70,159 training clips, while its AP is low. In contrast, the right column shows that ``Hoot'' only has 106 training clips, but achieves an AP of 0.86, and is larger than many other sound classes with more training clips. In the end of this article, we plot the mAP of all 527 sound classes in Fig. \ref{fig:long_fig}, which shows the class-wise comparison of the CNN14, MobileNetV1 and Wavegram-Logmel-CNN systems with previous state-of-the-art audio tagging system \cite{kong2019weakly} built with embedding features released by \cite{gemmeke2017audio}. The blue bars in Fig. \ref{fig:long_fig} show the number of training clips in logarithmic scale. \qk{The ``$+$'' symbol indicates label qualities between 0 and 1, which are measured by the percentage of correctly labelled audio clips verified by an expert \cite{gemmeke2017audio}. The label quality vary from sound class to sound class.} The ``$-$'' symbol indicates the sound classes whose label quality are not available. \qk{Fig. \ref{fig:long_fig} shows that the average precisions of some classes are higher than others. For example, sound classes such as ``bagpipes'' achieve an average precision of 0.90, while sound classes such as ``mouse'' achieve an average precision less than 0.2. One explanation is that the audio tagging difficulty is different between sound class to sound class. In addition, audio tagging performance is not always correlated with the number of training clips and label qualities \cite{kong2019weakly}.} Fig. \ref{fig:long_fig} shows that our proposed systems outperform previous state-of-the-art systems \cite{kong2018audio, yu2018multi} across a wide range of sound classes.  

\begin{table}[t]
\centering
\caption{Results with data balancing and augmentation}
\label{table:augmentation}
\begin{tabular}{lccc}
 \toprule
 Augmentation & mAP & AUC & d-prime \\
 \midrule
 no-bal,no-mixup (20k) & 0.224 & 0.894 & 1.763 \\
 bal,no-mixup (20k) & 0.221 & 0.879 & 1.652 \\
 bal,mixup (20k) & \textbf{0.278} & \textbf{0.905} & \textbf{1.850} \\
 \midrule
 no-bal,no-mixup (1.9m) & 0.375 & 0.971 & 2.690 \\
 bal,no-mixup (1.9m) & 0.416 & 0.968 & 2.613 \\
 bal,mixup (1.9m) & \textbf{0.431} & \textbf{0.973} & \textbf{2.732} \\
 bal,mixup-wav (1.9m) & 0.425 & 0.973 & 2.720 \\
 \bottomrule
\end{tabular}
\end{table}

\begin{table}[t]
\centering
\caption{Results of different hop sizes}
\label{table:hop_size}
\begin{tabular}{llccc}
 \toprule
 Hop size & Time resolution & mAP & AUC & d-prime \\
 \midrule
 1000 & 31.25 ms & 0.400 & 0.969 & 2.645 \\
 640 & 20.00 ms & 0.417 & 0.972 & 2.711 \\
 500 & 15.63 ms & 0.417 & 0.971 & 2.682 \\
 320 & 10.00 ms & \textbf{0.431} & \textbf{0.973} & \textbf{2.732} \\
 \bottomrule
\end{tabular}
\end{table}

\begin{table}[t]
\centering
\caption{Results of different embedding dimensions}
\label{table:embedding_dimensions}
\begin{tabular}{lccc}
 \toprule
 Embedding & mAP & AUC & d-prime \\
 \midrule
 32 & 0.364 & 0.958 & 2.437 \\
 128 & 0.412 & 0.969 & 2.634 \\
 512 & 0.420 & 0.971 & 2.689 \\
 2048 & \textbf{0.431} & \textbf{0.973} & \textbf{2.732} \\
 \bottomrule
\end{tabular}
\end{table}

\qk{
\begin{table}[t]
\centering
\caption{Results of partial training data}
\label{table:partial}
\begin{tabular}{lccc}
 \toprule
 Training data & mAP & AUC & d-prime \\
 \midrule
 50\% of full & 0.406 & 0.964 & 2.543 \\
 80\% of full & 0.426 & 0.971 & 2.677 \\
 100\% of full & \textbf{0.431} & \textbf{0.973} & \textbf{2.732} \\
 \bottomrule
\end{tabular}
\end{table}

\begin{table}[t]
\centering
\caption{Results of different sample rates}
\label{table:sample_rate}
\begin{tabular}{lccc}
 \toprule
 Sample rate & mAP & AUC & d-prime \\
 \midrule
 8 kHz & 0.406 & 0.970 & 2.654 \\
 16 kHz & 0.427 & 0.973 & 2.719 \\
 32 kHz & \textbf{0.431} & \textbf{0.973} & \textbf{2.732} \\
 \bottomrule
\end{tabular}
\end{table}

\begin{table}[t]
\centering
\caption{Results of different mel bins}
\label{table:mel_bins}
\begin{tabular}{lccc}
 \toprule
 Mel bins & mAP & AUC & d-prime \\
 \midrule
 32 bins & 0.413 & 0.971 & 2.691 \\
 64 bins & 0.431 & 0.973 & 2.732 \\
 128 bins & \textbf{0.442} & \textbf{0.973} & \textbf{2.735} \\
 \bottomrule
\end{tabular}
\end{table}
}

\qk{\subsubsection{Data balancing}
Section \ref{section:data_balancing} introduces the data balancing technique that we use to train AudioSet tagging systems. Fig. \ref{fig:six_figures}(b) shows the performance of the CNN14 system with and without data balancing. The blue curve shows that it takes a long time to train PANNs without data balancing. The green curve shows that with data balancing, a system converges faster within limited training iterations. In addition, the systems trained with full 1.9 million training clips perform better than the systems trained with the balanced subset of 20k training clips. Table \ref{table:augmentation} shows that the CNN14 system achieves an mAP of 0.416 with data balancing, higher than that without data balancing (0.375).

\subsubsection{Data augmentation}\label{section:exp_augmentation}
We show that mixup data augmentation plays an important role in training PANNs. By default, we apply mixup on the log mel spectrogram. Fig. \ref{fig:six_figures}(b) and Table \ref{table:augmentation} shows that the CNN14 system trained with mixup data augmentation achieves an mAP of 0.431, outperforming that trained without mixup data augmentation (0.416). Mixup is especially useful when training with the balanced subset containing only 20k training clips, yielding an mAP of 0.278, compared to training without mixup (0.221). In addition, we show that mixup on the log mel spectrogram achieves an mAP of 0.431, outperforming mixup in the time-domain waveform of 0.425, when training with full training data. This suggests that mixup is more effective when used with the log mel spectrogram than with the time-domain waveform.}

\subsubsection{Hop sizes}
The hop size is the number of samples between adjacent frames. A small hop size leads to high resolution in the time domain. We investigate the influence of different hop sizes on AudioSet tagging with the CNN14 system. We investigate hop sizes of 1000, 640, 500 and 320: these correspond to time domain resolutions of 31.25 ms, 20.00 ms, 15.63 ms and 10.00 ms between adjacent frames, respectively. Table \ref{table:hop_size} shows that the mAP score increases as hop sizes decrease. With a hop size of 320, the CNN14 system achieves an mAP of 0.431, outperforming the larger hop sizes of 500, 640 and 1000.

\subsubsection{Embedding dimensions}
Embedding features are fixed-length vectors that summarize audio clips. By default, the CNN14 has an embedding feature dimension of 2048. We investigate a rage of CNN14 systems with embedding dimensions of 32, 128, 512 and 2048. Fig. \ref{fig:six_figures}(c) and Table \ref{table:embedding_dimensions} show that mAP performance increases as embedding dimension increases.

\begin{table}[t]
\centering
\caption{Results of different systems}
\begin{tabular}{lccc}
 \toprule
 Architecture & mAP & AUC & d-prime \\
 \midrule
 CNN6 & 0.343 & 0.965 & 2.568 \\
 CNN10 & 0.380 & 0.971 & 2.678 \\
 CNN14 & \textbf{0.431} & \textbf{0.973} & \textbf{2.732} \\
 \midrule
 ResNet22 & 0.430 & 0.973 & 0.270 \\
 ResNet38 & \textbf{0.434} & \textbf{0.974} & \textbf{2.737} \\
 ResNet54 & 0.429 & 0.971 & 2.675 \\
 \midrule
 MobileNetV1 & \textbf{0.389} & \textbf{0.970} & \textbf{2.653} \\
 MobileNetV2 & 0.383 & 0.968 & 2.624 \\
 \midrule
 DaiNet \cite{dai2017very} & 0.295 & 0.958 & 2.437 \\
 LeeNet11 \cite{lee2017sample} & 0.266 & 0.953 & 2.371 \\
 LeeNet24 & 0.336 & 0.963 & 2.525 \\
 Res1dNet31 & \textbf{0.365} & \textbf{0.958} & \textbf{2.444} \\
 Res1dNet51 & 0.355 & 0.948 & 2.295 \\
 \midrule
 Wavegram-CNN & 0.389 & 0.968 & 2.612 \\
 Wavegram-Logmel-CNN & \textbf{0.439} & \textbf{0.973} & \textbf{2.720} \\
 \bottomrule
\end{tabular}
\label{table:all_systems}
\end{table}

\qk{\subsubsection{Training with partial data}
The audio clips of AudioSet are sourced from YouTube. Same audio clips are no longer downloadable, and others may be removed in the future. For better reproducibility of our work in future, we investigate the performance of systems trained with randomly chosen partial data ranging from 50\% to 100\% of our downloaded data. Fig. \ref{fig:six_figures}(d) and Table \ref{table:partial} show that the mAP decreases slightly from 0.431 to 0.426 (a 1.2\% drop) when the CNN14 system is trained with 80\% of full data, and decreases to 0.406 (a 5.8\% drop) when trained with 50\% of full data. This result shows that the amount of training data is important for training PANNs.

\subsubsection{Sample rate}
Fig. \ref{fig:six_figures}(e) and Table \ref{table:sample_rate} show the performance of the CNN14 system trained with different sample rate. The CNN14 system trained with 16 kHz audio recordings achieves an mAP of 0.427, similar (within 1.0\%) to the CNN14 system trained with a sample rate of 32 kHz. On the other hand, the CNN14 system trained with 8 kHz audio recordings achieves a lower mAP of 0.406 (5.8\% lower). This indicates that information in the 4 kHz - 8 kHz range is useful for audio tagging. 

\subsubsection{Mel bins}
Fig. \ref{fig:six_figures}(f) and Table \ref{table:mel_bins} show the performance of the CNN14 system trained with different number of mel bins. The system achieves an mAP of 0.413 with 32 mel bins, compared to 0.431 with 64 mel bins and 0.442 with 128 mel bins. This result suggests that PANNs achieve better performance with more mel bins, although the computation complexity increases linearly with the number of mel bins. Throughout this paper, we adopt 64 mel bins for extracting the log mel spectrogram, as a trade-off between computational complexity and system performance.}

\subsubsection{Number of CNN layers}
We investigate the performance of CNN systems with 6, 10 and 14 layers, as described in Section \ref{subsection:cnns}. Table \ref{table:all_systems} shows that the 6-, 10- and 14-layer CNNs achieve mAPs of 0.343, 0.380 and 0.431, respectively. This result suggests that PANNs with deeper CNN architectures achieve better performance than shallower CNN architectures. This result is in contrast to previous audio tagging systems trained on smaller datasets where shallower CNNs such as 9-layer CNN performed better than deeper CNNs \cite{kong2019cross}. One possible explanation is that smaller datasets may suffer from overfitting, while AudioSet is large enough to train deeper CNNs, at least up to the 14 layers CNNs that we investigate.

\subsubsection{ResNets}
We apply ResNets to investigate the performance of deeper PANNs. Table \ref{table:all_systems} shows that the ResNet22 system achieves an mAP of 0.430 similar to the CNN14 system. ResNet38 achieves an mAP of 0.434, slightly outperforming other systems. ResNet54 achieves an mAP of 0.429, which does not further improve the performance.

\subsubsection{MobileNets}

The systems mentioned above show that PANNs achieve good performance in AudioSet tagging. However, those systems do not consider computational efficiency when implemented on portable devices. We investigate building PANNs with light weight MobileNets described in Section \ref{section:mobile_net}. Table \ref{table:all_systems} shows that MobileNetV1 achieves an mAP of 0.389, 9.7\% lower to the CNN14 system of 0.431. The number of multiplication and addition (multi-adds) and parameters of the MobileNetV1 system are only 8.6\% and 5.9\% of the CNN14 system, respectively. The MobileNetV2 system achieves an mAP of 0.383, 11.1\% lower than CNN14, and is more computationally efficient than MobileNetV1, where the number of multi-adds and parameters are only 6.7\% and 5.0\% of the CNN14 system. 

\subsubsection{One-dimensional CNNs}
Table \ref{table:all_systems} shows the performance of one-dimensional CNNs. The DaiNet with 18 layers \cite{dai2017very} achieves an mAP of 0.295. The LeeNet11 with 11 layers \cite{lee2017sample} achieves an mAP of 0.266. Our improved LeeNet with 24 layers improves the mAP of LeeNet11 to 0.336. Our proposed Res1dNet31 and Res1dNet51 described in Section \ref{section:one_dim_cnns} achieve mAPs of 0.365 and 0.355 respectively, and achieve state-of-the-art performance among one-dimensional CNN systems.

\subsubsection{Wavegram-Logmel-CNN}
The bottom rows of Table \ref{table:all_systems} show the result of our proposed Wavegram-CNN and Wavegram-Logmel-CNN systems. The Wavegram-CNN system achieves an mAP of 0.389, outperforming the best previous one-dimensional CNN system (Res1dNet31). This indicates that the Wavegram is an effective learnt feature. Furthermore, our proposed Wavegram-Logmel-CNN system achieves a state-of-the-art mAP of 0.439 among all PANNs.

\begin{table}[t]
\centering
\caption{Number of multi-adds and parameters of different systems}
\label{table:complexity}
\begin{tabular}{{lrr}}
 \toprule
 Architecture & Multi-Adds & Parameters \\
 \midrule
 CNN6 & 21.986 $\times 10^9$ & 4,837,455 \\
 CNN10 & 28.166 $\times 10^9$ & 5,219,279 \\
 CNN14 & 42.220 $\times 10^9$ & 80,753,615 \\
 ResNet22 & 30.081 $\times 10^9$ & 63,675,087\\
 ResNet38 & 48.962 $\times 10^9$ & 73,783,247 \\
 ResNet54 & 54.563 $\times 10^9$ & 104,318,159 \\
 MobileNetV1 & 3.614 $\times 10^9$ & 4,796,303 \\
 MobileNetV2 & 2.810 $\times 10^9$ & 4,075,343 \\
 DaiNet & 30.395 $\times 10^9$ & 4,385,807 \\
 LeeNet11 & 4.741 $\times 10^9$ & 748,367 \\
 LeeNet24 & 26.369 $\times 10^9$ & 10,003,791 \\
 Res1dNet31 & 32.688 $\times 10^9$ & 80,464,463 \\
 Res1dNet51 & 61.833 $\times 10^9$ & 106,538,063 \\
 Wavegram-CNN & 44.234 $\times 10^9$ & 80,991,759 \\
 Wavegram-Logmel-CNN & 53.510 $\times 10^9$ & 81,065,487 \\
 \bottomrule
\end{tabular}
\end{table}

\begin{figure}[t]
  \centering
  \centerline{\includegraphics[width=0.8\columnwidth]{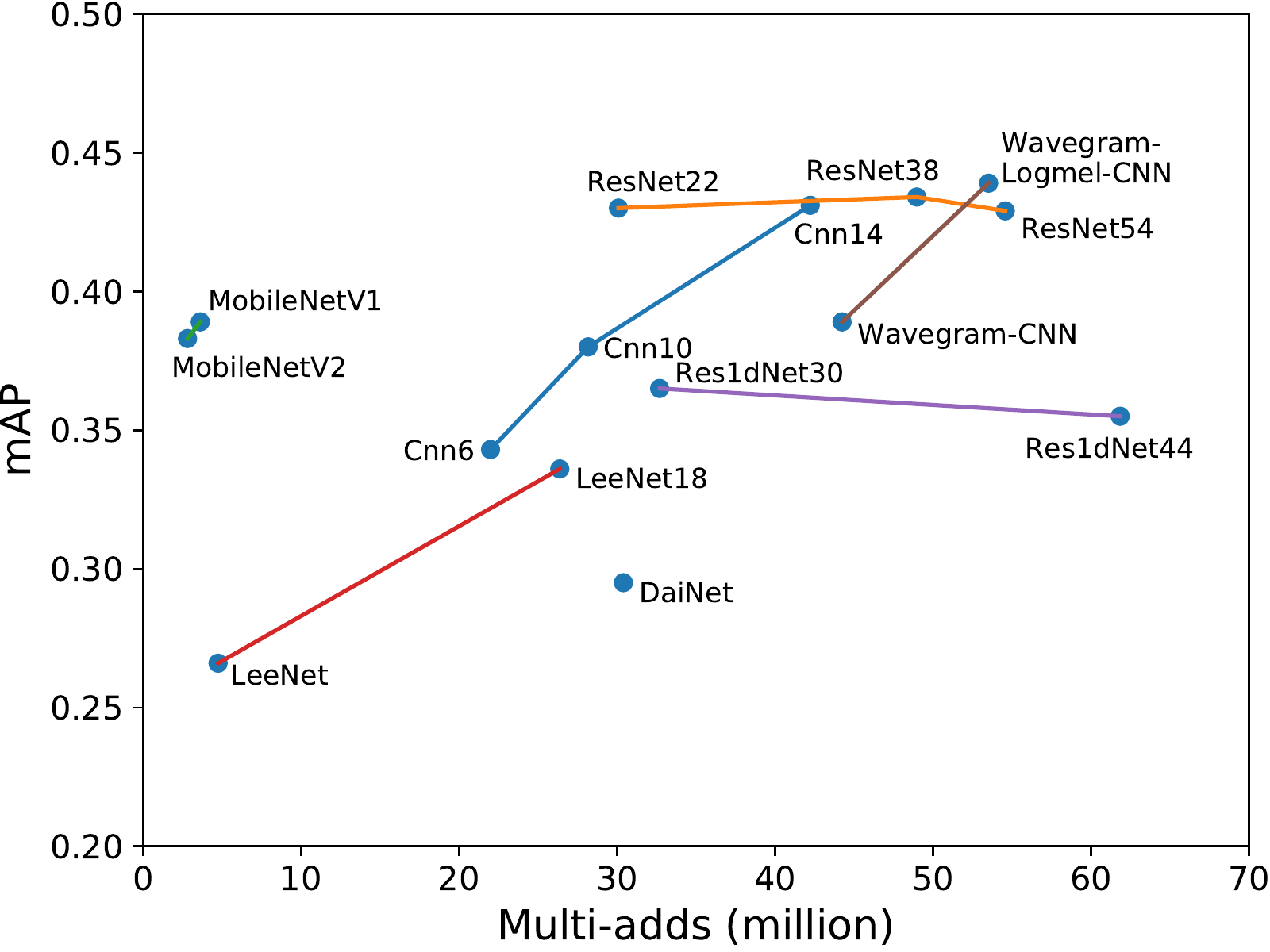}}
  \caption{Multi-adds versus mAP of AudioSet tagging systems. The same types of architectures are grouped in the same color.}
  \label{fig:complexity_mAP}
\end{figure}

\subsubsection{Complexity analysis}
We analyze the computational complexity of PANNs for inference. The number of multi-adds and parameters are two important factors for complexity analysis. The middle column of Table \ref{table:complexity} shows the number of multi-adds to infer a 10-second audio clip. The right column of Table \ref{table:complexity} shows the number of parameters of different systems. The number of multi-adds and parameters of the CNN14 system are $42.2 \times 10^9$ and 80.8 million, respectively, which are larger than the CNN6 and CNN10 systems. The number of multi-adds for the ResNets22 and ResNet38 systems are slightly less than for the CNN14 system. The ResNet54 system contains the most multi-adds at $54.6 \times 10^9$. One-dimensional CNNs have a similar computational cost to the two-dimensional CNNs. The best performing one-dimensional system, Res1dNet31, contains $32.7 \times 10^9$ multi-adds and 80.5 million parameters. Our proposed Wavegram-CNN system contains $44.2 \times 10^9$ multi-adds and 81.0 million parameters, which is similar to CNN14. The Wavegram-Logmel-CNN system slightly increases the multi-adds to $53.5 \times 10^9$, and the number of parameters is to 81.1 million, which is similar to CNN14. To reduce the number of multi-adds and parameters, MobileNets are applied. The MobileNetV1 and MobileNetV2 systems are light weight CNNs, with only $3.6 \times 10^9$ and $2.8 \times 10^9$ multi-adds and around 4.8 million and 4.1 million parameters, respectively. MobileNets reduce both the computational cost and system size. Figure \ref{fig:complexity_mAP} summarizes the mAP versus multi-adds of different PANNs. The same type of systems are linked by lines of the same color. The mAP increases from bottom to top. On the top-right is our proposed Wavegram-Logmel-CNN system that achieves the best mAP. On the top-left are MobileNetV1 and MobileNetV2 that are the most computationally efficient systems.

\subsection{Transfer to other tasks}
In this section, we investigate the application of PANNs to a range of other pattern recognition tasks. PANNs can useful for few-shot learning, for the tasks where only a limited number of training clips are provided. Few-shot learning is an important research topic in audio pattern recognition, as collecting labelled data can be time consuming. We transfer PANNs to other audio pattern recognition tasks using the methods described in Section \ref{section:transfer}. To begin with, we resample all audio recordings to 32 kHz, and convert them to monophonic to be consistent with the PANNs trained on AudioSet. We perform the following strategies described in Section \ref{section:transfer} for each task: 1) Train a system from scratch; 2) Use a PANN as a feature extractor; 3) Fine-tune a PANN. When using a PANN as the feature extractor, we build classifiers on the embedding features with one and three fully-connected layers, and call them ``Freeze + 1 layers'' (Freeze\_L1) and ``Freeze + 3 layers'' (Freeze\_L3), respectively. \qk{We adopt the CNN14 system for transfer learning to provide a fair comparison with other CNN based systems for audio pattern recognition.} We also investigate the performance of PANNs trained with different number of shots when training other audio pattern recognition tasks.

\begin{figure}[t]
  \centering
  \centerline{\includegraphics[width=0.9\columnwidth]{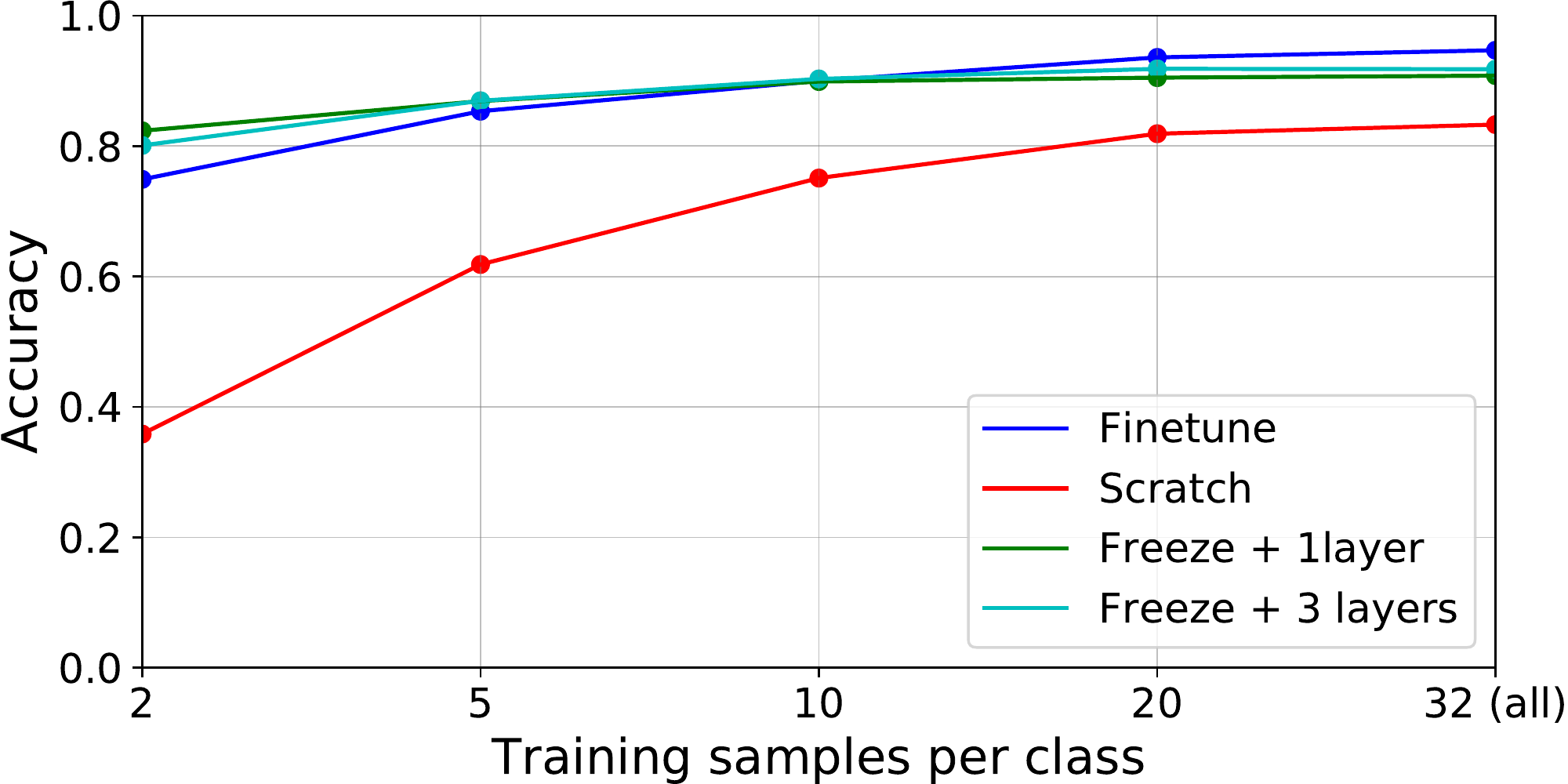}}
  \caption{Accuracy of ESC-50 with various number of training clips per class.}
  \label{fig:esc50}
\end{figure}

\begin{table}[t]
\centering
\caption{Accuracy of ESC-50}
\label{table:esc50}
\begin{tabular}{*{6}{c}}
 \toprule
 & STOA \cite{sailor2017unsupervised} & Scratch & fine-tune & Freeze\_L1 & Freeze\_L3 \\
 \midrule
 Acc. & 0.865 & 0.833 & \textbf{0.947} & 0.908 & 0.918 \\
 \bottomrule
\end{tabular}
\end{table}

\subsubsection{ESC-50}
ESC-50 is an environmental sound dataset \cite{piczak2015esc} consisting of 50 sound events, such as ``Dog'' and ``Rain''. There are 2,000 5-second audio clips in the dataset, with 40 clips per class. Table \ref{table:esc50} shows the 5-fold cross validation \cite{piczak2015esc} accuracy of the CNN14 system. Sailor et al. \cite{sailor2017unsupervised} proposed a state-of-the-art system for ESC-50, achieved an accuracy of 0.865 using unsupervised filterbank learning with a convolutional restricted Boltzmann machine. Our fine-tuned system achieves an accuracy of 0.947, outperforming previous state-of-the-art system by a large margin. The Freeze\_L1 and Freeze\_L3 systems achieve accuracies of 0.918 and 0.908, respectively. Training the CNN14 system from scratch achieves an accuracy of 0.833. Fig. \ref{fig:esc50} shows the accuracy of ESC-50 with different numbers of training clips of each sound class. Using a PANN as a feature extractor achieves the best performance when fewer than 10 clips per sound class are available for training. With more training clips, the fine-tuned systems achieve better performance. Both the fine-tuned system and the system using the PANN as a feature extractor outperform the systems trained from scratch. 

\begin{figure}[t]
  \centering
  \centerline{\includegraphics[width=0.9\columnwidth]{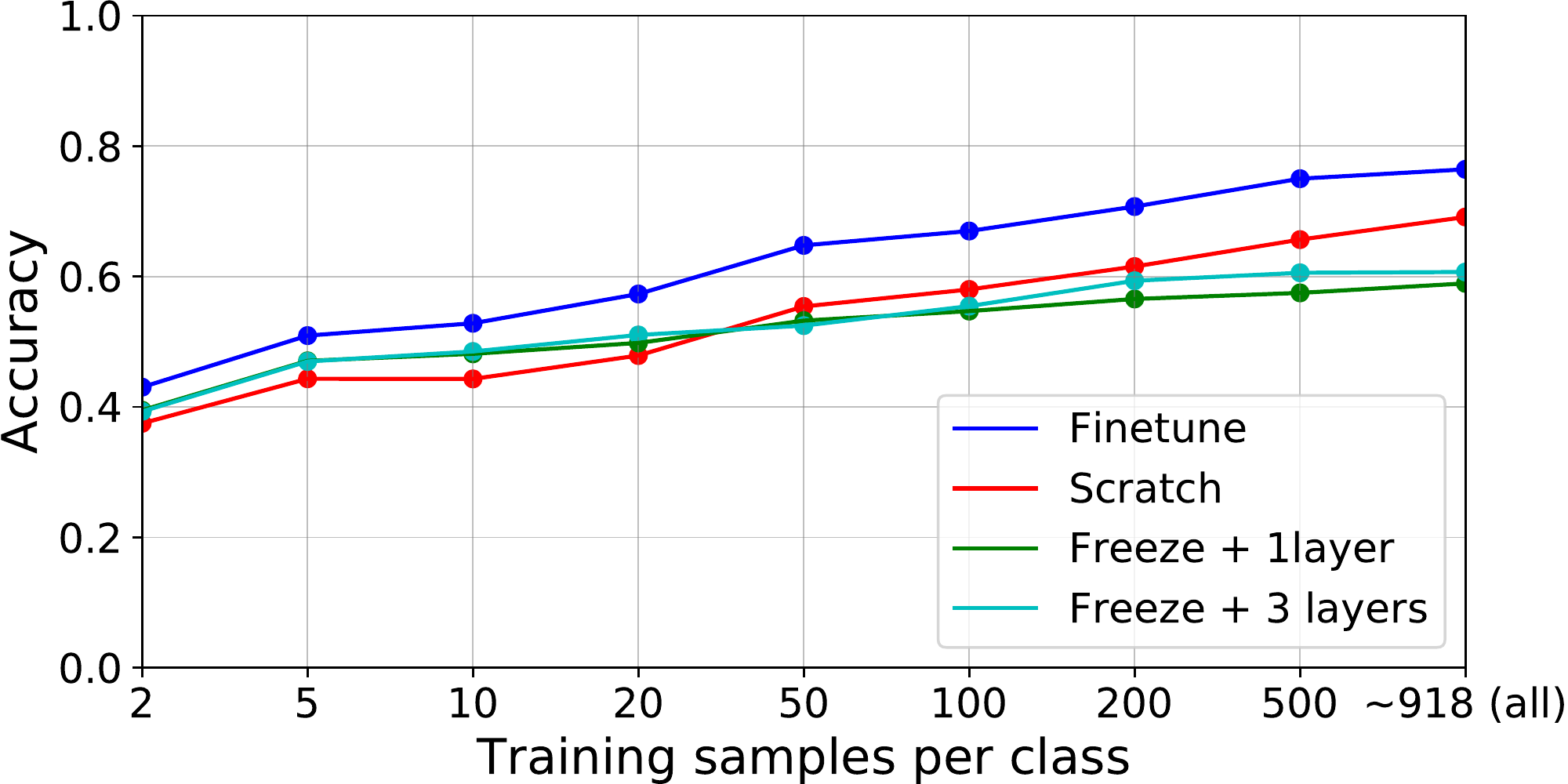}}
  \caption{Accuracy of DCASE 2019 Task 1 with various number of training clips per class.}
  \label{fig:dcase2019_task1}
\end{figure}

\begin{table}[t]
\centering
\caption{Accuracy of DCASE 2019 Task 1}
\label{table:dcase2019_task1}
\begin{tabular}{*{6}{c}}
 \toprule
 & STOA \cite{chen2019integrating} & Scratch & Fine-tune & Freeze\_L1 & Freeze\_L3 \\
 \midrule
 Acc. & \textbf{0.851} & 0.691 & 0.764 & 0.589 & 0.607 \\
 \bottomrule
\end{tabular}
\end{table}

\begin{figure}[t!]
  \centering
  \centerline{\includegraphics[width=0.9\columnwidth]{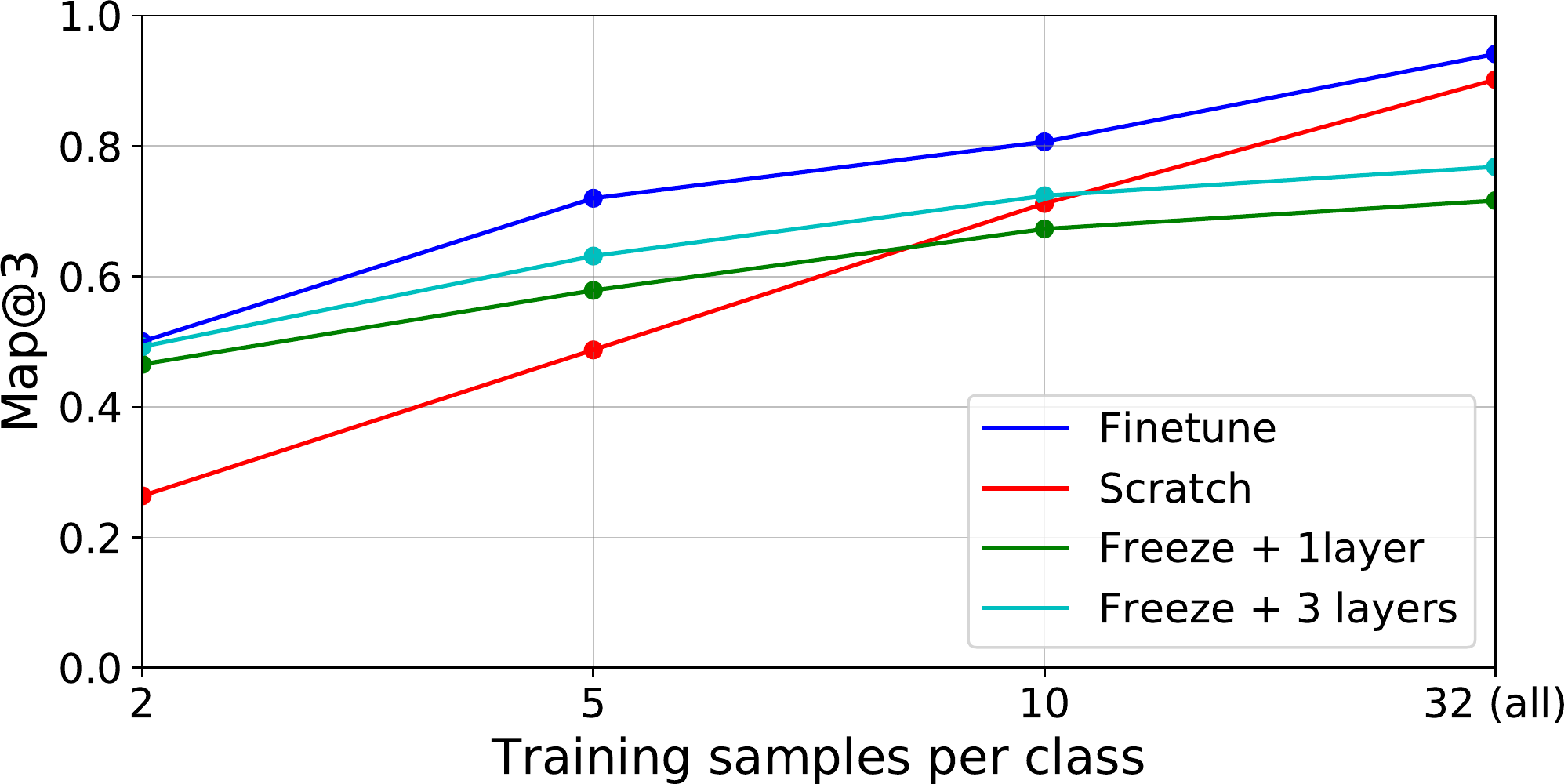}}
  \caption{Accuracy of DCASE 2018 Task 2 with various number of training clips per class.}
  \label{fig:dcase2018_task2}
\end{figure}

\begin{table}[t!]
\centering
\caption{Accuracy of DCASE 2018 Task 2}
\label{table:dcase2018_task2}
\begin{tabular}{*{6}{c}}
 \toprule
 & STOA \cite{jeong2018audio} & Scratch & Fine-tune & Freeze\_L1 & Freeze\_L3 \\
 \midrule
 mAP@3 & \textbf{0.954} & 0.902 & 0.941 & 0.717 & 0.768 \\
 \bottomrule
\end{tabular}
\end{table}

\begin{figure}[t]
  \centering
  \centerline{\includegraphics[width=0.9\columnwidth]{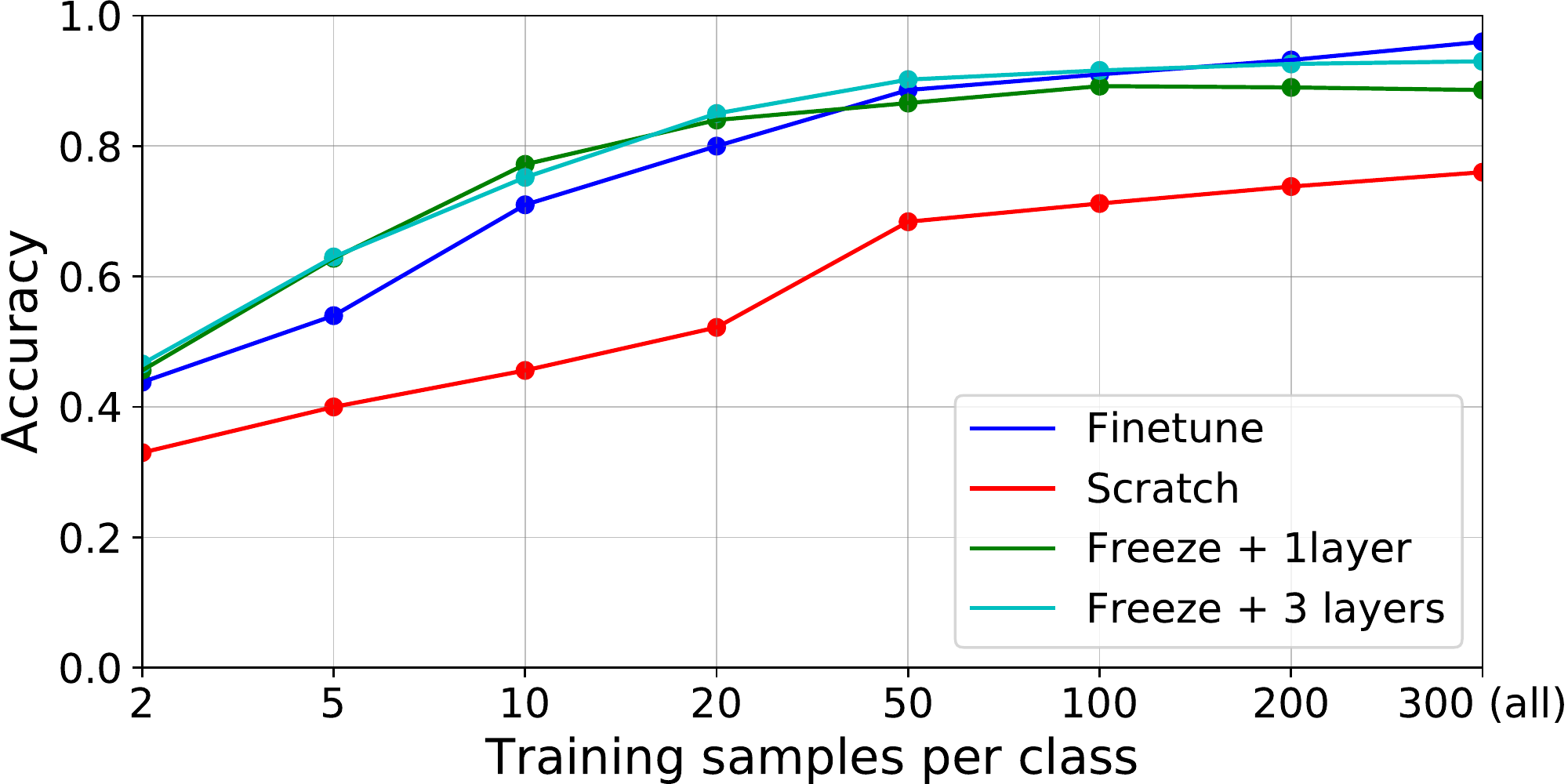}}
  \caption{Accuracy of MSoS with various number of training clips per class.}
  \label{fig:msos}
\end{figure}

\begin{table}[t]
\centering
\caption{Accuracy of MSoS}
\label{table:msos}
\begin{tabular}{*{6}{c}}
 \toprule
 & STOA \cite{chenattention} & Scratch & Fine-tune & Freeze\_L1 & Freeze\_L3 \\
 \midrule
 Acc. & 0.930 & 0.760 & \textbf{0.960} & 0.886 & 0.930 \\
 \bottomrule
\end{tabular}
\end{table}

\subsubsection{DCASE 2019 Task 1}
DCASE 2019 Task 1 is an acoustic scene classification task \cite{mesaros2018multi}, with a dataset consisting of over 40 hours of stereo recordings collected from various acoustic scenes in 12 European cities. We focus on Subtask A, where each audio recording has two channels with a sampling rate of 48 kHz. In the development set, there are 9185 and 4185 audio clips for training and validation respectively. We convert the stereo recordings to monophonic by averaging the stereo channels. CNN14 trained from scratch achieves an accuracy of 0.691, while the fine-tuned system achieves an accuracy of 0.764. Freeze\_L1 and Freeze\_L3 achieve accuracies of 0.689 and 0.607 respectively, and do not outperform the CNN14 trained from scratch. One possible explanation for this underperformance is that the audio recordings of acoustic scene classification have different distribution of AudioSet. So using PANN as a feature extractor does not outperform fine-tune or train a system from scratch. The fine-tuned system achieves better performance than the system trained from scratch. Fig. \ref{fig:dcase2019_task1} shows the classification accuracy of systems with various numbers of training clips per class. Table \ref{table:dcase2019_task1} shows the overall performance. The state-of-the-art system of Chen et al \cite{chen2019integrating}. achieves an accuracy of 0.851 using the combination of various classifiers and stereo recordings as input, while we do not use any ensemble methods and stereo recordings.

\subsubsection{DCASE 2018 Task 2}
DCASE 2018 Task 2 is a general-purpose automatic audio tagging task \cite{Fonseca2018_DCASE} using a dataset of audio recordings from Freesound annotated with a vocabulary of 41 labels from the AudioSet ontology. The development set consists of 9,473 audio recordings with durations from 300 ms to 30 s. The mAP@3 is used for evaluating system performance \cite{Fonseca2018_DCASE}. \qk{In training, we break or pad audio recordings into 4-second audio segments. In inference, we average the predictions of those segments to obtain the prediction of an audio recording.} Table \ref{table:dcase2018_task2} shows that the best previous method proposed by Jeong and Lim \cite{jeong2018audio} achieves an mAP@3 of 0.954 using an ensemble of several systems. In comparison, our CNN14 system trained from scratch achieves an accuracy of 0.902. The fine-tuned CNN14 system achieves an mAP@3 of 0.941. The Freeze\_L1 and Freeze\_L3 systems achieve accuracies of 0.717 and 0.768 respectively. Fig. \ref{fig:dcase2018_task2} shows the mAP@3 with different numbers of training clips. The fine-tuned CNN14 system outperforms the systems trained from scratch and the systems using PANN as a feature extractor. The fine-tuned CNN14 system achieves comparable results to the state-of-the-art system.

\subsubsection{MSoS}
The Making Sense of Sounds (MSoS) data challenge \cite{kroos2019generalisation} is a task to predict an audio recording to one of five categories: ``Nature'', ``Music'', ``Human'', ``Effects'' and ``Urban''. The dataset consists of a development set of 1,500 audio clips and an evaluation set of 500 audio clips. All audio clips have a duration of 4 seconds. The state-of-the-art system proposed by Chen and Gupta \cite{chenattention} achieves an accuracy of 0.930. Our fine-tuned CNN14 achieves an accuracy of 0.960, outperforming previous state-of-the-art system. CNN14 trained from scratch achieves an accuracy of 0.760. Fig. \ref{fig:msos} shows the accuracy of the systems with different number of training clips. The fine-tuned CNN14 system and the system using CNN14 as a feature extractor outperforms CNN14 trained from scratch. 

\begin{figure}[t]
  \centering
  \centerline{\includegraphics[width=0.9\columnwidth]{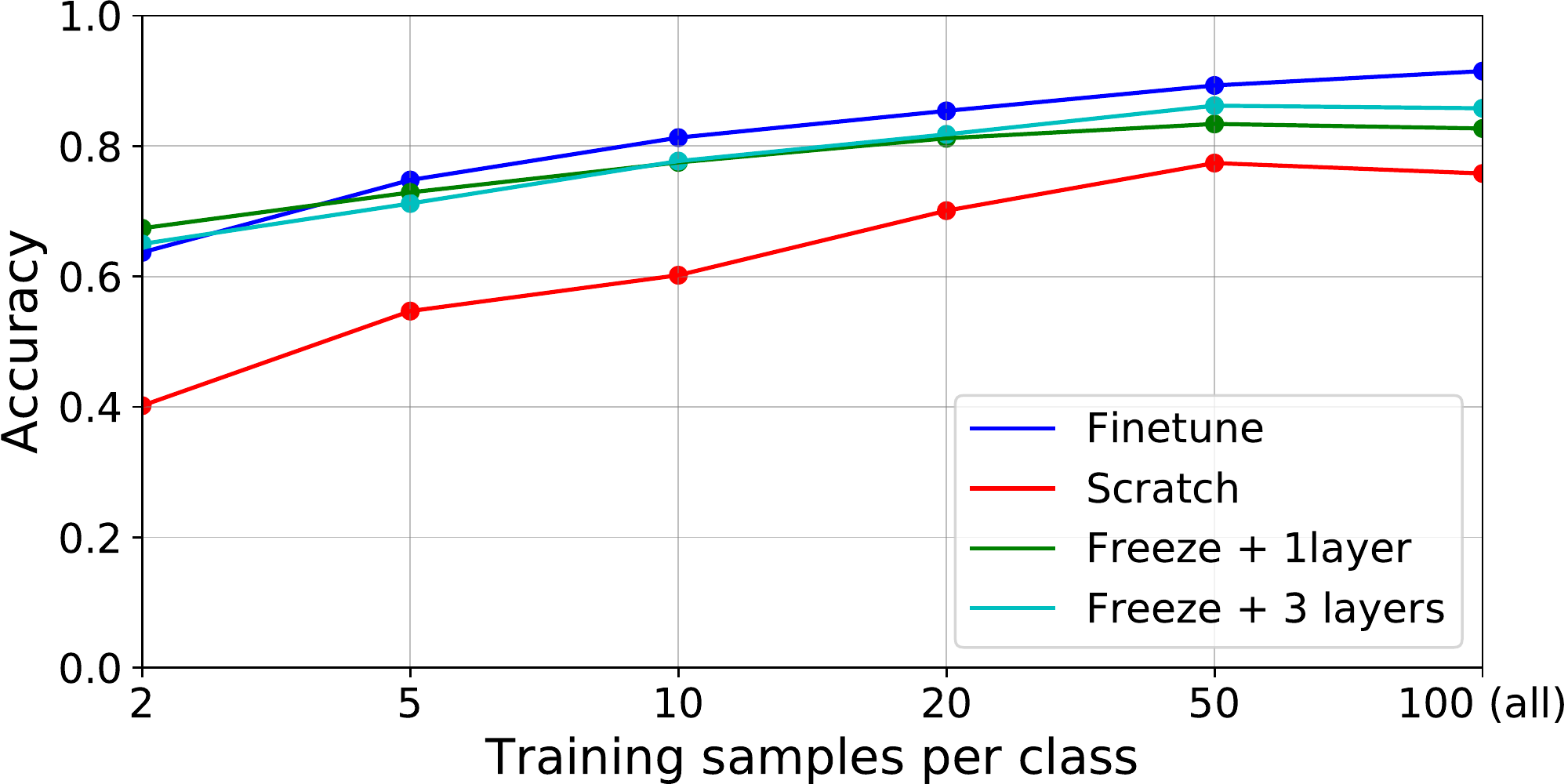}}
  \caption{Accuracy of GTZAN with various number of training clips per class.}
  \label{fig:gtzan}
\end{figure}

\begin{table}[t]
\centering
\caption{Accuracy of GTZAN}
\label{table:gtzan}
\begin{tabular}{*{6}{c}}
 \toprule
 & STOA \cite{liu2019bottom} & Scratch & Fine-tune & Freeze\_L1 & Freeze\_L3 \\
 \midrule
 Acc. & \textbf{0.939} & 0.758 & 0.915 & 0.827 & 0.858 \\
 \bottomrule
\end{tabular}
\end{table}

\subsubsection{GTZAN}
The GTZAN dataset \cite{tzanetakis2002musical} is a music genre classification dataset containing 1,000 30-second music clips of 10 genres of music, such as ``Classical'' and ``Country''. All music clips have a duration of 30 seconds and a sampling rate of 22,050 Hz. In development, 10-fold cross validation is used to evaluate the performance of systems. Table \ref{table:gtzan} shows that previous state-of-the-art system proposed by Liu et al. \cite{liu2019bottom} achieves an accuracy of 0.939 using a bottom-up broadcast neural network. The fine-tuned CNN14 system achieves an accuracy of 0.915, outperforming CNN14 trained from scratch with an accuracy of 0.758 and the Freeze\_L1 and Freeze\_L3 systems with accuracies of 0.827 and 0.858 respectively. Fig. \ref{fig:gtzan} shows the accuracy of systems with different numbers of training clips. The Freeze\_L1 and Freeze\_L3 systems achieve better performance than other systems trained with less than 10 clips per class. With more training clips, the fine-tuned CNN14 system performs better than other systems. 

\begin{figure}[t!]
  \centering
  \centerline{\includegraphics[width=0.9\columnwidth]{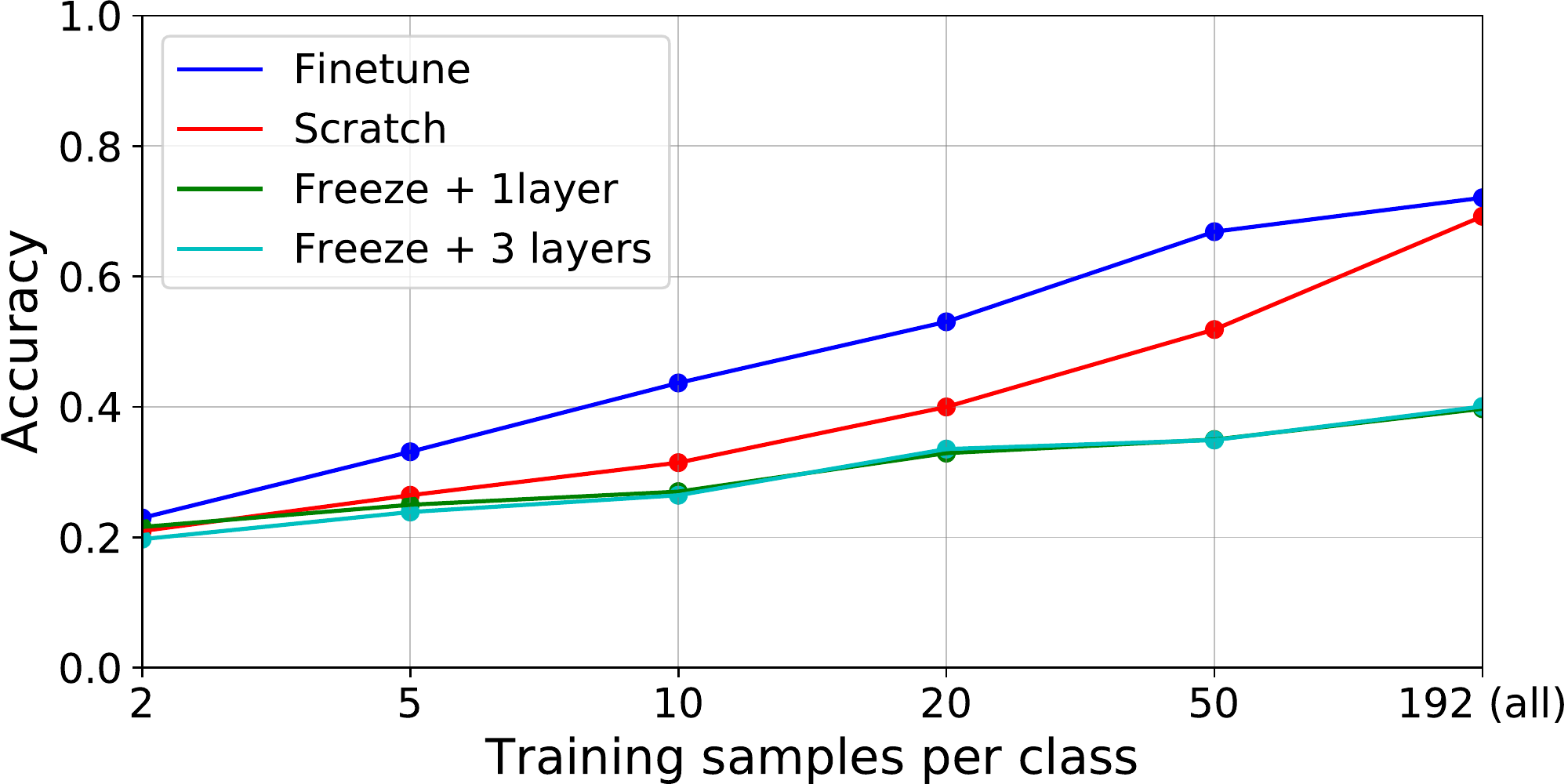}}
  \caption{Accuracy of RAVDESS with various number of training clips per class.}
  \label{fig:ravdess}
\end{figure}

\begin{table}[t!]
\centering
\caption{Accuracy of RAVDESS}
\label{table:ravdness}
\begin{tabular}{*{6}{c}}
 \toprule
 & STOA \cite{zeng2019spectrogram} & Scratch & Fine-tune & Freeze\_L1 & Freeze\_L3 \\
 \midrule
 Acc. & 0.645 & 0.692 & \textbf{0.721} & 0.397 & 0.401 \\
 \bottomrule
\end{tabular}
\end{table}

\subsubsection{RAVDESS}
The Ryerson Audio-Visual Database of Emotional Speech and Song (RAVDESS) is a human speech emotion dataset \cite{livingstone2012ravdess}. The database consists of sounds from 24 professional actors including 12 female and 12 male simulating 8 emotions, such as ``Happy'' and ``Sad''. The task is to classify each sound clip into an emotion. There are 1,440 audio clips in the development set. We evaluate our systems with 4-fold cross validation. Table \ref{table:ravdness} shows that previous state-of-the-art system proposed by Zeng et al. \cite{zeng2019spectrogram} achieves an accuracy of 0.645. Our CNN14 system trained from scratch achieves an accuracy of 0.692. The fine-tuned CNN14 system achieves a state-of-the-art accuracy of 0.721. The Freeze\_L1 and Freeze\_L3 systems achieve much lower accuracies of 0.397 and 0.401 respectively. Fig. \ref{fig:ravdess} shows the accuracy of systems with respect to a range of training clips. The fine-tuned systems and the system trained from scratch outperform the system using PANN as a feature extractor. This suggests that audio recordings of the RAVDESS dataset may have different distributions of the AudioSet dataset. Therefore, the parameters of a PANN need be fine-tuned to achieve good performance on the RAVDESS classification task.

\subsection{Discussion}
In this article, we have investigated a wide range of PANNs for AudioSet tagging. Several of our proposed PANNs have outperformed previous state-of-the-art AudioSet tagging systems, including CNN14 achieves an mAP of 0.431, and ResNet38 achieves an mAP of 0.434, outperforming Google's baseline of 0.314. MobileNets are light-weight systems that have fewer multi-adds and numbers of parameters. MobileNetV1 achieves an mAP of 0.389. Our adapted one-dimensional system Res1dNet31 achieves an mAP of 0.365, outperforming previous one-dimensional CNNs including DaiNet \cite{dai2017very} of 0.295 and LeeNet11 \cite{lee2017sample} of 0.266. Our proposed Wavegram-Logmel-CNN system achieves the highest mAP of 0.439 among all PANNs. \qk{PANNs can be used as a pretrained model for new audio pattern recognition tasks.}

PANNs trained on the AudioSet dataset were transferred to six audio pattern recognition tasks. We show that fine-tuned PANNs achieve state-of-the-art performance in the ESC-50, MSOS and RAVDESS classification tasks, and approach the state-of-the-art performance in the DCASE 2018 Task 2 and the GTZAN classification task. Of the PANN systems, the fine-tuned PANNs always outperform PANNs trained from scratch on new tasks. The experiments show that PANNs have been successful in generalizing to other audio pattern recognition tasks with limited number of training data.

\section{Conclusion}\label{section:conclusion}
We have presented pretrained audio neural networks (PANNs) trained on the AudioSet for audio pattern recognition. A wide range of neural networks are investigated to build PANNs. We propose a Wavegram feature learnt from waveform, and a Wavegram-Logmel-CNN that achieves state-of-the-art performance in AudioSet tagging, archiving an mAP of 0.439. We also investigate the computational complexity of PANNs. We show that PANNs can be transferred to a wide range of audio pattern recognition tasks and outperform several previous state-of-the-art systems. PANNs can be useful when fine-tuned on a small amount of data on new tasks. In the future, we will extend PANNs to more audio pattern recognition tasks. 
% Can use something like this to put references on a page
% by themselves when using endfloat and the captionsoff option.

\ifCLASSOPTIONcaptionsoff
  \newpage
\fi

% trigger a \newpage just before the given reference
% number - used to balance the columns on the last page
% adjust value as needed - may need to be readjusted if
% the document is modified later
%\IEEEtriggeratref{8}
% The "triggered" command can be changed if desired:
%\IEEEtriggercmd{\enlargethispage{-5in}}

% references section

% can use a bibliography generated by BibTeX as a .bbl file
% BibTeX documentation can be easily obtained at:
% http://mirror.ctan.org/biblio/bibtex/contrib/doc/
% The IEEEtran BibTeX style support page is at:
% http://www.michaelshell.org/tex/ieeetran/bibtex/
%\bibliographystyle{IEEEtran}
% argument is your BibTeX string definitions and bibliography database(s)
%\bibliography{IEEEabrv,../bib/paper}
%
% <OR> manually copy in the resultant .bbl file
% set second argument of \begin to the number of references
% (used to reserve space for the reference number labels box)
%\begin{thebibliography}{1}
%\bibitem{IEEEhowto:kopka}
%H.~Kopka and P.~W. Daly, \emph{A Guide to \LaTeX}, 3rd~ed.\hskip 1em plus
%  0.5em minus 0.4em\relax Harlow, England: Addison-Wesley, 1999.
%\end{thebibliography}

\bibliographystyle{IEEEtran}
\bibliography{refs}

% biography section
% 
% If you have an EPS/PDF photo (graphicx package needed) extra braces are
% needed around the contents of the optional argument to biography to prevent
% the LaTeX parser from getting confused when it sees the complicated
% \includegraphics command within an optional argument. (You could create
% your own custom macro containing the \includegraphics command to make things
% simpler here.)
%\begin{IEEEbiography}[{\includegraphics[width=1in,height=1.25in,clip,keepaspectratio]{mshell}}]{Michael Shell}
% or if you just want to reserve a space for a photo:

% if you will not have a photo at all:

% You can push biographies down or up by placing
% a \vfill before or after them. The appropriate
% use of \vfill depends on what kind of text is
% on the last page and whether or not the columns
% are being equalized.

%\vfill

% Can be used to pull up biographies so that the bottom of the last one
% is flush with the other column.
%\enlargethispage{-5in}

\begin{IEEEbiography}[{\includegraphics[width=1in,height=1.25in,clip,keepaspectratio]{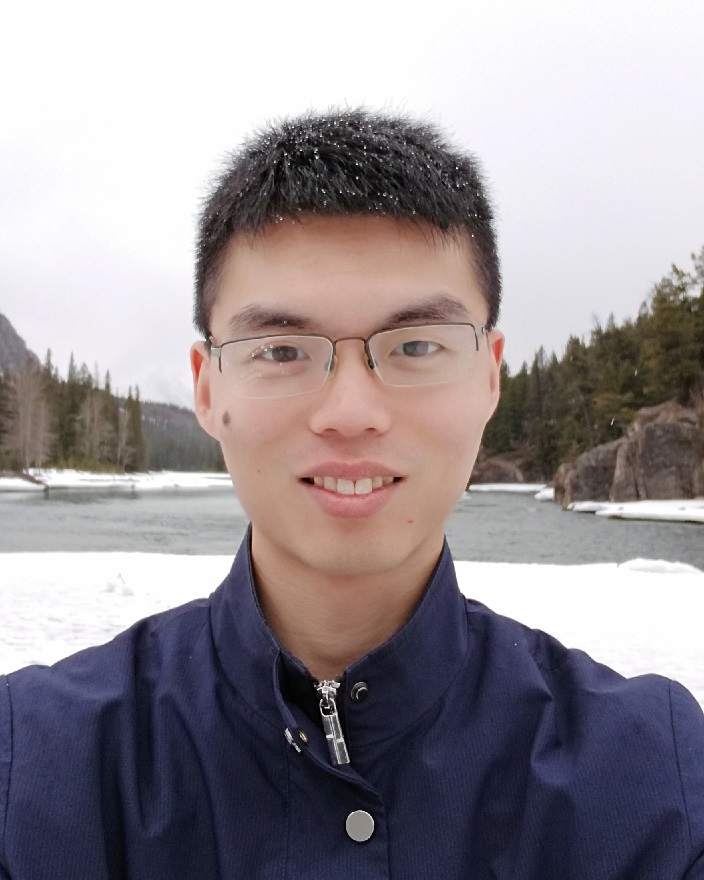}}]{Qiuqiang Kong}
(S'17) received the B.Sc. degree in 2012, and the M.E. degree in 2015 from South China University of Technology, Guangzhou, China. He received the Ph.D. degree from University of Surrey, Guildford, UK in 2020. Following his PhD, he joined ByteDance AI Lab as a research scientist. His research topic includes the classification, detection and separation of general sounds and music. He is known for developing attention neural networks for audio tagging, and winning the audio tagging task in the detection and classification of acoustic scenes and events (DCASE) challenge in 2017. He was nominated as the postgraduate research student of the year in University of Surrey, 2019. He is a frequent reviewer for journals and conferences in the area including IEEE/ACM Transactions on Audio, Speech, and Language Processing.
\end{IEEEbiography}

\vskip 0pt plus -1fil

\begin{IEEEbiography}[{\includegraphics[width=1in,height=1.25in,clip,keepaspectratio]{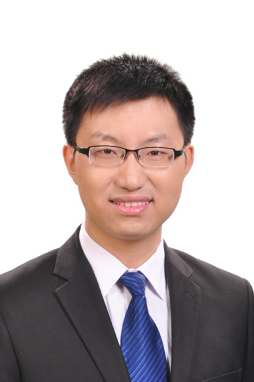}}]{Yin Cao} (M'18) received the B.Sc. degree in Electronic Science and Engineering from Nanjing University, China in 2008, and Ph.D. degree from Institute of Acoustics, Chinese Academy of Sciences, China in 2013. He then worked in Acoustics group at Brigham Young University, US, and at Institute of Acoustics, Chinese Academy of Sciences, China. In 2018, he joined Centre for Vision, Speech and Signal Processing at the University of Surrey. His research topic includes active noise control, air acoustics and signal processing, detection, classification and separation of audio. He is known for his work on decentralized active noise control, weighted spatial gradients control metric, and polyphonic sound event detection and localization. He was the winner of urban sound tagging in detection and classification of acoustic scenes and events (DCASE) 2020 challenge and achieved second-best of sound event detection and localization tasks in DCASE 2019 challenge. He has served as an Associate Editor for Noise Control Engineering Journal since 2020. He is also a frequent reviewer for IEEE/ACM Transactions on Audio, Speech, and Language Processing.
\end{IEEEbiography}

\vskip 0pt plus -1fil

\begin{IEEEbiography}[{\includegraphics[width=1in,height=1.25in,clip,keepaspectratio]{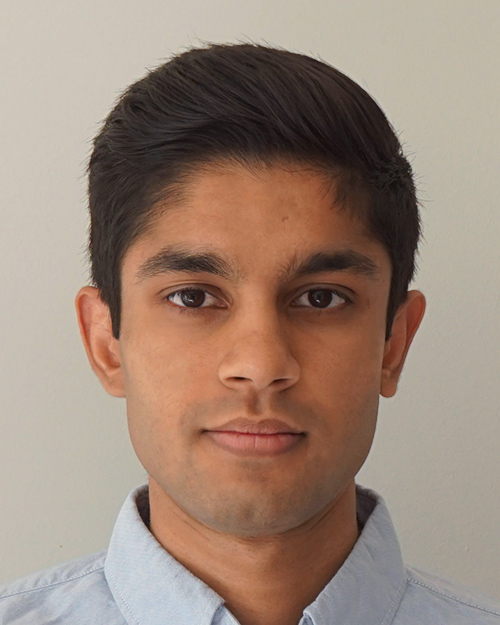}}]{Turab Iqbal} received the B.Eng. degree in Electronic Engineering from the University of Surrey, U.K., in 2017. Currently, he is working towards a Ph.D. degree from the Centre for Vision, Speech and Signal Processing (CVSSP) in the University of Surrey. During his time as a Ph.D. student, he has worked on a number of projects in the area of audio classification and localization using machine learning methods. His research is mainly focused on learning with weakly-labeled or noisy training data.
\end{IEEEbiography}

\vskip 0pt plus -1fil

\vskip 0pt plus -1fil

\begin{IEEEbiography}[{\includegraphics[width=1in,height=1.25in,clip,keepaspectratio]{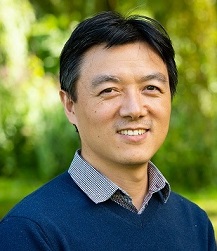}}]{Wenwu Wang} (M'02-SM'11) was born in Anhui, China. He received the B.Sc. degree in 1997, the M.E. degree in 2000, and the Ph.D. degree in 2002, all from Harbin Engineering University, China. He then worked in King's College London, Cardiff University, Tao Group Ltd. (now Antix Labs Ltd.), and Creative Labs, before joining University of Surrey, UK, in May 2007, where he is currently a professor in signal processing and machine learning, and a Co-Director of the Machine Audition Lab within the Centre for Vision Speech and Signal Processing. He has been a Guest Professor at Qingdao University of Science and Technology, China, since 2018. His current research interests include blind signal processing, sparse signal processing, audio-visual signal processing, machine learning and perception, machine audition (listening), and statistical anomaly detection. He has (co)-authored over 200 publications in these areas. He served as an Associate Editor for IEEE Transactions on Signal Processing from 2014 to 2018. He is also Publication Co-Chair for ICASSP 2019, Brighton, UK. He currently serves as Senior Area Editor for IEEE Transactions on Signal Processing and an Associate Editor for IEEE/ACM Transactions on Audio Speech and Language Processing.
\end{IEEEbiography}

\vskip 0pt plus -1fil

\begin{IEEEbiography}[{\includegraphics[width=1in,height=1.25in,clip,keepaspectratio]{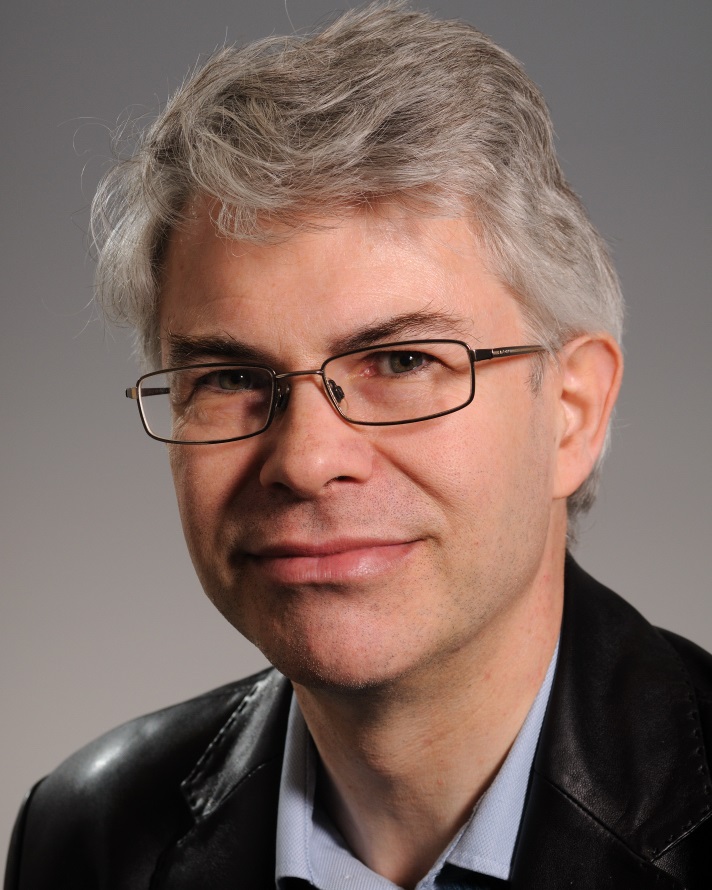}}]{Mark D. Plumbley}
(S'88-M'90-SM'12-F'15) received the B.A.(Hons.) degree in electrical sciences and the Ph.D. degree in neural networks from University of Cambridge, Cambridge, U.K., in 1984 and 1991, respectively. Following his PhD, he became a Lecturer at King's College London, before moving to Queen Mary University of London in 2002. He subsequently became Professor and Director of the Centre for Digital Music, before joining the University of Surrey in 2015 as Professor of Signal Processing. He is known for his work on analysis and processing of audio and music, using a wide range of signal processing techniques, including matrix factorization, sparse representations, and deep learning. He is a co-editor of the recent book on 
Computational Analysis of Sound Scenes and Events,
and Co-Chair of the recent DCASE 2018 Workshop on Detection and Classifications of Acoustic Scenes and Events. He is a Member of the IEEE Signal Processing Society Technical Committee on Signal Processing Theory and Methods, and a Fellow of the IET and IEEE.

\end{IEEEbiography}

% that's all folks
\end{document}